\documentclass[twoside,11pt]{article}
\usepackage{graphicx}
\usepackage{natbib}
\usepackage{amsmath}
\usepackage{lipsum} 
\usepackage{floatrow}
\newfloatcommand{capbtabbox}{table}[][\FBwidth]

\usepackage[sc]{mathpazo} 
\usepackage[T1]{fontenc} 
\usepackage{microtype} 

\usepackage[a4paper,hmarginratio=1:1,top=25mm,columnsep=20pt,textwidth=185mm]{geometry} 
\usepackage{multicol} 
\usepackage{hyperref} 

\usepackage[small,labelfont=bf,up,textfont=it,up]{caption} 
\usepackage{booktabs} 
\usepackage{float} 

\usepackage{paralist} 

\usepackage{abstract} 

\usepackage{titlesec} 
\renewcommand\thesection{\Roman{section}}
\titleformat{\section}[block]{\large\scshape\centering}{\thesection.}{1em}{} 

\usepackage{fancyhdr} 
\usepackage[dvipsnames*,svgnames]{xcolor}
\usepackage[framemethod=default]{mdframed}
\mdfdefinestyle{NR}{skipabove=\topskip,skipbelow=\topskip,%
                    ,align=center,%
                    innerleftmargin=.25cm,linecolor=black,%
                    linewidth=0pt,backgroundcolor=Gainsboro}
\newmdenv[style=NR]{NR}
\mdfdefinestyle{NR1}{align=center,%
                     innerleftmargin=.25cm,linecolor=black,%
                     linewidth=0.5pt,backgroundcolor=Lavender}
\newmdenv[style=NR1]{NR1}

\newcommand{\simlt}{\lower.5ex\hbox{$\; \buildrel < \over \sim \;$}}
\newcommand{\simgt}{\lower.5ex\hbox{$\; \buildrel > \over \sim \;$}}
\newcommand{\etal}{{\it et al.}}
\def\HII{\hbox{H$\scriptstyle\rm II\ $}}

\def\HI{\hbox{H$\scriptstyle\rm I\ $}}

\usepackage{tcolorbox}
\usepackage{tabularx}
\usepackage{array}
\usepackage{colortbl}
\tcbuselibrary{skins}

\newcolumntype{Y}{>{\raggedleft\arraybackslash}X}
\newcolumntype{C}[1]{>{\centering\let\newline\\\arraybackslash\hspace{0pt}}m{#1}}

\tcbset{tab1/.style={enhanced,fonttitle=\bfseries,fontupper=\normalsize\sffamily,
colback=yellow!10!white,colframe=red!50!black,colbacktitle=Salmon!40!white,
coltitle=black,center title}}

\title{\vspace{-4mm}
\textbf{\fontsize{20pt}{10pt}\selectfont
{\tt CHRONOS\footnote{Chronos, the Greek god personifying time is
the name we have given to this survey, as it is designed to 
understand the formation and evolution of galaxies across cosmic time}}\\\vspace{+2mm} 
\fontsize{16pt}{10pt}\selectfont
A NIR Spectroscopic Galaxy Survey:\\
\vspace{-1mm}
\fontsize{15pt}{10pt}\selectfont
{\bf From the formation of galaxies to the peak of activity}}\\
{\normalsize --- A proposal sent in response to ESA's Voyage 2050 call ---}
\vspace{+2mm}}

\author{\large
  {\bf Lead Proposer}\\
  \textsc{Ignacio Ferreras$^{1,2}$}\thanks{email: i.ferreras@ucl.ac.uk}
\vspace{+1mm}\\
{\bf Science Core Team}\\
\textsc{Mark Cropper$^3$, Ray Sharples$^4$,}\\
\vspace{-4mm}\\
\textsc{Joss Bland-Hawthorn$^5$, Gustavo Bruzual$^6$, St\'ephane Charlot$^7$, Christopher J. Conselice$^8$,}\\ 
\textsc{Simon Driver$^9$, James Dunlop$^{10}$, Andrew~M. Hopkins$^{11}$, Sugata Kaviraj$^{12}$,}\\
\textsc{Tom Kitching$^3$, Francesco La Barbera$^{13}$, Ofer Lahav$^2$, Anna Pasquali$^{14}$,}\\
\textsc{Stephen Serjeant$^{15}$, Joseph Silk$^{7,16,17}$, Rogier Windhorst$^{18}$}\\
}
\date{Submitted August 5$^{\rm th}$, 2019}
\vspace{-5mm}

\begin{document}

\setcounter{page}{1}
\maketitle 
\thispagestyle{empty}

\noindent
{\small $^1$ Instituto de Astrof\'\i sica de Canarias, E-38200 La Laguna, Tenerife, Spain\\
$^2$ Department of Physics and Astronomy, University College London, Gower Street, London WC1E 6BT, UK\\
$^3$ Mullard Space Science Laboratory, University College London, Dorking, Surrey RH5 6NT, UK\\
$^4$ Department of Physics, Durham University, South Road, Durham DH1 3LE, UK\\
$^5$ Sydney Institute for Astronomy, School of Physics, University of Sydney, NSW 2006, Australia\\
$^6$ Instituto de Radioastronom\'ia y Astrof\'isica, UNAM, Campus Morelia, 58089 Morelia, M\'exico\\
$^7$ UPMC-CNRS, UMR7095, Institut d\'{}Astrophysique de Paris, F-75014 Paris, France\\
$^8$ School of Physics \& Astronomy, University of Nottingham, Nottingham NG7 2RD, UK\\
$^9$ ICRAR, M468, University of Western Australia, 35 Stirling Hwy, Crawley, WA 6009, Australia\\
$^{10}$ SUPA, Institute for Astronomy, University of Edinburgh, Royal Observatory, Edinburgh EH9 3HJ, UK\\
$^{11}$ Australian Astronomical Optics, Macquarie University, 105 Delhi Rd, North Ryde, NSW 2113, Australia\\
$^{12}$ Centre for Astrophysics Research, University of Hertfordshire, College Lane, Hatfield AL10 9AB, UK\\
$^{13}$ INAF-Osservatorio Astronomico di Capodimonte, sal. Moiariello 16, I-80131 Napoli, Italy\\
$^{14}$ Astronomisches Rechen-Institut/ZAH, Universit\"at Heidelberg, M\"onchhofstr. 12-14 69120 Heidelberg, Germany\\
$^{15}$ School of Physical Sciences, The Open University, Milton Keynes, MK7 6AA, UK\\
$^{16}$ Dept. of Physics and Astronomy, The Johns Hopkins University Homewood Campus, Baltimore, MD 21218, USA\\
$^{17}$ Sub-department of Astrophysics, University of Oxford, Keble Road, Oxford OX1 3RH, UK\\
$^{18}$ School of Earth and Space Exploration, Arizona State University, Tempe, AZ 85287, USA
}

\clearpage

\thispagestyle{fancy} 
\phantom{.}
\vspace{+2mm}

\setcounter{page}{1}
\begin{NR}[userdefinedwidth=165mm]
\begin{center}
{\bf\Large Executive Summary}
\end{center}

\vskip+2mm

Responding to ESA's call regarding the Voyage 2050 long-term plan to define the
future space missions that will address the astrophysics science questions during the
2035-2050 cycle, we propose a dedicated, ultra-deep spectroscopic survey in
the near infrared (NIR), that will target a mass-limited sample of
galaxies during two of the most fundamental epochs of cosmic
evolution: the formation of the first galaxies
(at z$\simgt 6$; {\sl cosmic dawn}), and at the peak of galaxy formation
activity (at redshift z$\sim$1--3; {\sl cosmic noon}). By way of NIR
observations ($\lambda$=0.8--2$\mu$m), it is possible to study the
UV\,Lyman-$\alpha$ region in the former, and the optical rest-frame in
the latter, allowing us to extract fundamental observables such as gas
and stellar kinematics, chemical abundances, and ages, providing a
unique legacy database covering these two crucial stages of cosmic
evolution.
\bigskip

The faintness of these distant sources and the need for a large number
of spectra -- to produce robust statistical constraints on the
mechanisms of galaxy evolution -- pose one of the strongest challenges faced
by any astrophysics mission to date. To put the mission in context,
the survey requires the equivalent of gathering one Hubble Ultra-Deep
Field\footnote{\tt http://www.spacetelescope.org/science/deep\_fields/}
every fortnight for five years. Furthermore, the need to work in the
NIR at extremely low flux levels makes a ground-based approach
unfeasible due to atmospheric emission and absorption.  Only with the
largest facilities of the future (e.g. {\sl ELT}) will be
possible to observe a reduced set of targets, comprising at most of order 
1,000s of galaxies. Likewise, from space, the small field of view of
{\sl JWST} and its use as a general purpose facility will yield a
rather small set of high quality NIR spectra of distant galaxies
(in the thousands, at best). Our
project (codename {\tt Chronos}) aims to produce $\sim$1\,million
high quality spectra, with a high S/N in the {\sl continuum}, where
information about the underlying stellar populations is encoded. We note
that cosmology-driven redshift surveys impose much weaker constraints
on the S/N in the continuum, as they only use galaxies as ``test
particles'', thus only requiring a redshift measurement.
This project focuses on the galaxies themselves. The
proposed database is needed to solve the key open questions in galaxy
formation. More specifically, the main science drivers are:\smallskip
\begin{compactitem}
\item The connection between the star formation history and the mass assembly history.
\item The role of AGN and supernova feedback in shaping the 
  formation histories of galaxies, with a quantitative estimate of quenching timescales.
\item The formation of the first galaxies.
\item The source of reionization.
\item Evolution of the metallicity-mass relation, including [$\alpha$/Fe] and
  individual abundances.
\item Precision cosmology through detailed studies of the ``baryon physics'' of
  galaxy formation, probing the power spectrum over scales k$\sim$1\,Mpc$^{-1}$.
\end{compactitem}
\smallskip
The purpose of this proposal is to start a comprehensive study of such
a demanding survey, focusing on the challenging technical aspects
involving an ultra-deep (H$\sim$24-26\,AB), high multiplex ($\simgt$5,000),
NIR (0.8--2\,$\mu$m) space-based spectrograph, at optimal resolution for
galaxy formation studies ($\sim$2,000), and with a large
field of view ($\simgt$0.2\,deg$^2$).

\vskip+1mm
\phantom{.}
\end{NR}

\clearpage

%

\section{The next steps in extragalatic astrophysics}
\label{sec:Intro}
\vskip+0.5truecm
\begin{multicols}{2} 
The era of extragalactic astrophysics began in earnest around the
time of the Great Debate between Shapley and Curtis in
1920 \citep{Trimble:95}. The debate focused on the nature of a number
of intriguing ``nebulae'', at a time when the consensus rested on a Universe
in which the Milky Way was its main constituent,
a scenario that harks back to the model of our Galaxy laid out by
\citet{Herschel:1785}. The discoveries
during the 1930s, pioneered by Slipher, Hubble, and
Humason resulted in the concept of island Universes,
where each ``stellar system'', a galaxy, constitutes a
fundamental building block tracing the largest scales in the
Cosmos. It has been nearly a century since this Great Debate, and our
understanding of extragalactic astrophysics has come a long
way.
\bigskip

Developments in telescopes, instrumentation and analysis techniques
have allowed us to decipher the intricacies of galaxy formation. At
present, the established paradigm rests on a dark matter dominated
cosmic web within which a comparatively small mass fraction consists
of ordinary matter (``baryons''), mostly in the form of stars, gas and
dust. The first stage of galaxy formation is driven by the 
(linear) growth of the dark matter density fluctuations imprinted
during the earliest phases of cosmic evolution. 
Stable dark matter structures, termed halos, collapse and virialise,
constituting the basic units in this scenario. At the same time, 
gas accumulates in the central regions of these
halos, leading to cooling and star formation. The general aspects of
this complex process can be explained within the current framework
\citep[see, e.g.][]{SM:12}, resulting in an overall very successful
theory that matches the observations. 
However, many of the key processes are
only roughly understood, most notably the ``baryon physics'' that
transforms the smooth distribution of gas at early times into the
galaxies we see today. This complex problem requires large, targeted 
data sets probing the most important phases of galaxy
formation and evolution.  This proposal addresses the next steps that
the astrophysics community will follow in the near future to
understand structure formation. High-quality spectroscopic
observations of galaxies are required to probe these important
phases. {\sl We motivate below the need for a large, space-based, ultra-deep survey of
galaxy spectra in the near-infrared, and present the technological
challenges that must be addressed.}

The extremely weak fluxes of the targets, combined with the need to work
at near infrared wavelengths imply such a task must be pursued from space,
free of the noise from atmospheric emission and absorption. Moreover,  the
need to simultaneously observe many sources spectroscopically, from an
unmanned, unserviceable mission, defines arguably one of the toughest challenges
in space science. Such a task is optimally suited for the 2035-2050 period
envisioned by ESA within the Voyage 2050 call.  We emphasize that this science case
complements the succesful track record of ESA in this field,
with missions such as Herschel (tracing the evolution of dust in galaxies),
Gaia (tracing the gravitational potential of our Galaxy), as well as 
the cosmology-orientated missions, {\sl Planck} and {\sl Euclid}. We will show below how
the fundamental science case of galaxy formation and evolution requires
a future space-based observatory, beyond the capabilities of the
upcoming {\sl JWST} or large 30-40m ground-based telescopes such as ESO's {\sl ELT}.

\end{multicols}

\section{The evolution of galaxies at the peak of activity}
\label{sec:D4000}
\begin{multicols}{2} 

\subsection{Star formation across cosmic time}

The observational evidence reveals that the overall level of star
formation in nearby galaxies is comparatively low with respect to
earlier epochs. Fig.~\ref{fig:CSFH}, derived from various
observational traces of star formation, shows a characteristic peak in
the cosmic star formation activity in galaxies between redshifts 1 and
3, roughly corresponding to a cosmic time between 2 and 6\,Gyr after
the Big Bang (or a lookback time between 8 and 12\,Gyr ago).  Such a
trend can be expected as the gas from the initial stages is gradually
locked into stars and, subsequently, remnants. This trend is highly
packed with complex information

\begingroup
\centering
\includegraphics[width=85mm]{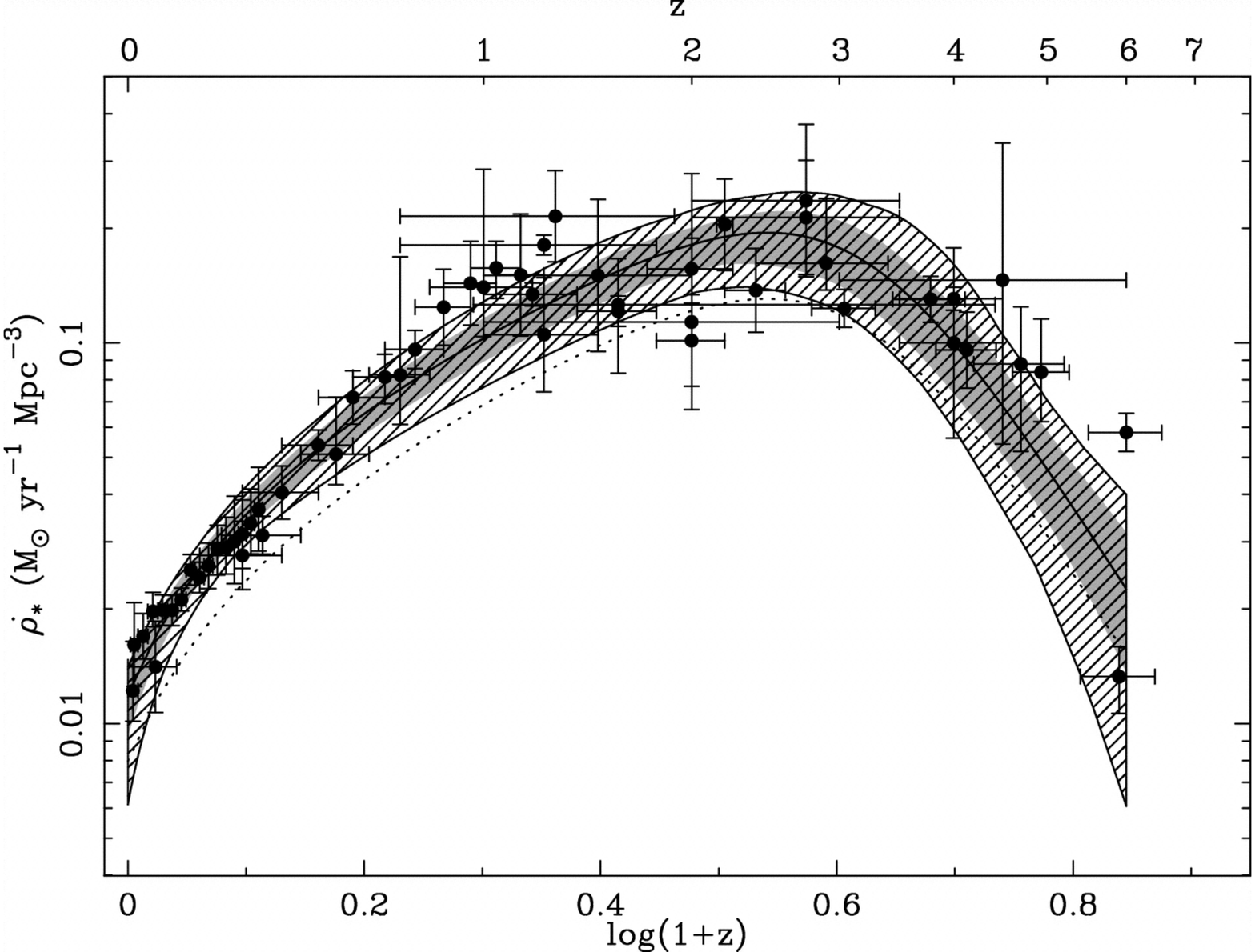}
\captionof{figure}{Cosmic star formation history: This diagram shows
  the redshift evolution of the
  star formation rate density.  Note that detailed
  spectroscopic {\sl optical} galaxy surveys exist only out to
  z$\simlt$1, whereas the epochs of maximum star formation 
  (z$\sim$1--3), and the first stages of formation (z$\simgt$7)
  are poorly understood (from \citealt{CSFH_HB06}; see also \citealt{CSFH_MD14})}
\label{fig:CSFH}
\vskip+0.3truecm
\endgroup

\noindent
regarding the efficiency of star
formation, the mechanisms of gas infall and outflows, the ejection
of gas from evolved phases of stellar evolution and the bottom-up
hierarchy of structure formation.

In addition, the z$\sim$1--3
redshift window corresponds to the peak of AGN activity \citep{Richards:06},
and merger rate \citep{Ryan:08}. 
Moreover, it is the epoch
when the dark matter halos hosting massive galaxies allow for
cold accretion via cosmic streams (see \S\S\ref{SS:ColdAcc}).
Decoding this complex puzzle
requires a detailed study of the different phases of evolution. At
present we only have complete galaxy samples amounting to $\sim$1 million
high quality spectra at low redshift (z$\simlt 0.2$,
e.g. SDSS, \citealt{SDSS}), along with samples of spectra at
intermediate redshift (z$\simlt$1.5), e.g.  VIPERS \citep{vipers},
VVDS \citep{vvds}, zCOSMOS \citep{zcosmos}, GAMA \citep{gama}, BOSS
\citep{boss} or LEGA-C \citep{LegaC}. Future spectroscopic surveys
will also probe similar redshift ranges within the optical spectral
window  -- e.g., WAVES \citep{WAVES}; WEAVE \citep{WEAVE}; DESI \citep{DESI},
MSE \citep{MSE}. In the NIR, ESO's VLT/MOONS \citep{MOONS} will constitute
the state-of-the art ground based survey, but the expected S/N will not
be high enough for studies comparable to those perfomed on SDSS spectra
at z$\simlt$0.2. We note that many of the spectroscopic surveys (past and future) 
are mostly designed as a ``redshift machine'' (i.e. optimised for cosmology, using
galaxies simply as ``test particles''), and the S/N of the data in the
continuum is too low for any of the science presented here to be
successfully delivered.  {\sl None of the current and future observing
  facilities, both ground- and space-based, will be capable of
  creating the equivalent of the spectroscopic SDSS catalogue at these
  redshifts.}

\subsection{Bimodality and galaxy assembly}

On a stellar mass vs colour (or age) diagram, galaxies populate two well
defined regions: the red sequence and the blue cloud \citep[see,
  e.g.,][]{Kauff:03,Taylor:15}. Galaxies on the red sequence are mostly massive,
passively-evolving systems with little or no ongoing star
formation. Although the red sequence extends over a wide range in
stellar mass, the most massive galaxies tend to be on the red
sequence, with a preferential early-type morphology. In contrast, blue
cloud galaxies have substantial ongoing star formation, and extend
towards the low-mass end.  A third component is also defined, the
green valley \citep{M07}, between these two. However, the identification of this
region as a transition stage between the blue cloud and the red
sequence is far from trivial \citep{Schaw:14,Ang:19}. There are many studies
tracing the redshift evolution of galaxies in these regions
\citep[e.g.][]{Bell:04,Ilbert:10,Muzz:13}, revealing a downsizing
trend, so that the bulk of star formation (i.e. the ``weight'' of the
blue cloud) shifts from the most massive galaxies at high redshift, to
lower mass systems in the present epoch.  This simple diagram allows
us to present a simplified version of star formation in galaxies,
including the usual bottom-up hierarchy that begins with small star
forming systems, leading to more massive galaxies through in situ star
formation and mergers, both with (``wet'') and without (``dry'')
additional star formation.

Models such as those proposed by
\citet[][see Fig.~\ref{fig:BiMod}]{Faber:07} allow us to express
graphically the complex processes involved. However, the problem with
these analyses is how to properly characterize the formation stage of
a galaxy by a simple descriptor such as galaxy colour. More detailed
analyses have been presented of the colour-mass diagram, combining
photometry and spectroscopy in relatively nearby samples
\citep[e.g.][]{Schaw:07} showing interesting processes that relate the
various sources of feedback (see \S\S\ref{SS:feedback}
below).

\begingroup \centering
\includegraphics[width=60mm]{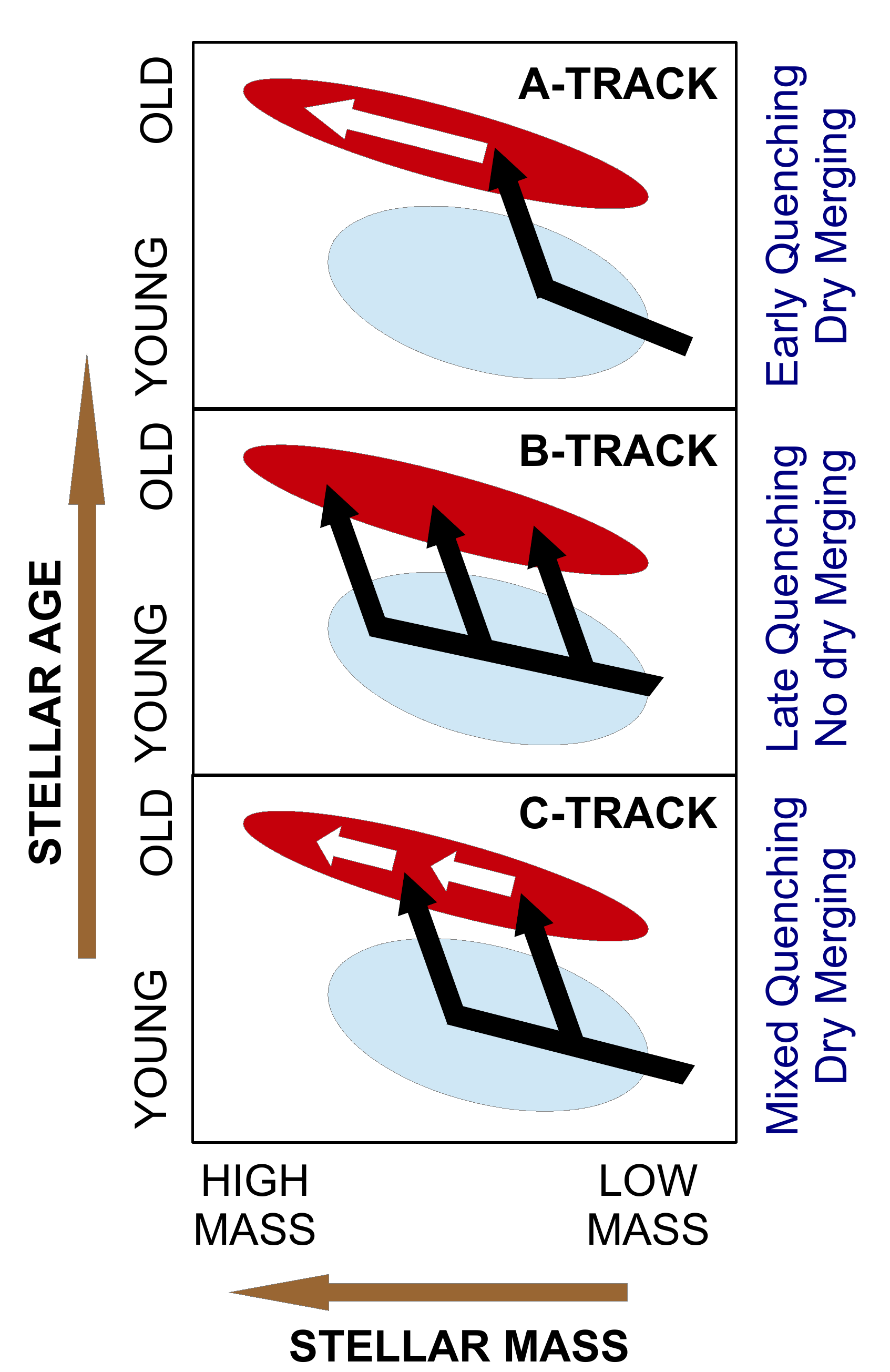}
\captionof{figure}{Schematics of galaxy evolution from the blue cloud
  to the red sequence. Three different scenarios are considered,
  as labelled, with the black arrows representing evolution 
  through wet mergers and quenching, and white arrows symbolising
  stellar mass growth through dry mergers \citep[adapted from][]{Faber:07}}
\label{fig:BiMod}
\vskip+0.3truecm
\endgroup

However, such studies are complicated by the fact that the
underlying stellar populations span a wide range of ages and chemical
composition, and the star formation processes do not involve a
substantial fraction of the baryonic mass of the galaxy.  Therefore,
it is necessary to extend these studies, including high quality
spectroscopic data, to explore the evolution on the colour-stellar mass
diagram with galaxies targeted during the peak of galaxy formation.
At these redshifts (z$\sim$1-3), we will be dealing with the most
important stages of formation.

\subsection{The role of star formation and AGN}
\label{SS:feedback}

The bimodality plot (Fig.~\ref{fig:BiMod}) illustrates the key 
processes underlying galaxy evolution. Most importantly, the presence
of a large population of passive galaxies on the red sequence, without
an equivalent counterpart of massive galaxies on the blue cloud requires
physical mechanisms by which star formation is quenched. As the fuel
for star formation is cold gas, quenching of any type must resort to
reducing this component, either by heating, photoionisation or mechanical
removal of the cold phase.

Various theoretical models have been explored over the
past decades, most notably based on the expulsion of
gas from supernovae-driven winds \citep[stellar feedback, e.g.][]{DekelSilk:86}
or from a central supermassive black
hole \citep[AGN feedback, e.g.][]{SilkRees:98}.

\begingroup
\centering
\includegraphics[width=80mm]{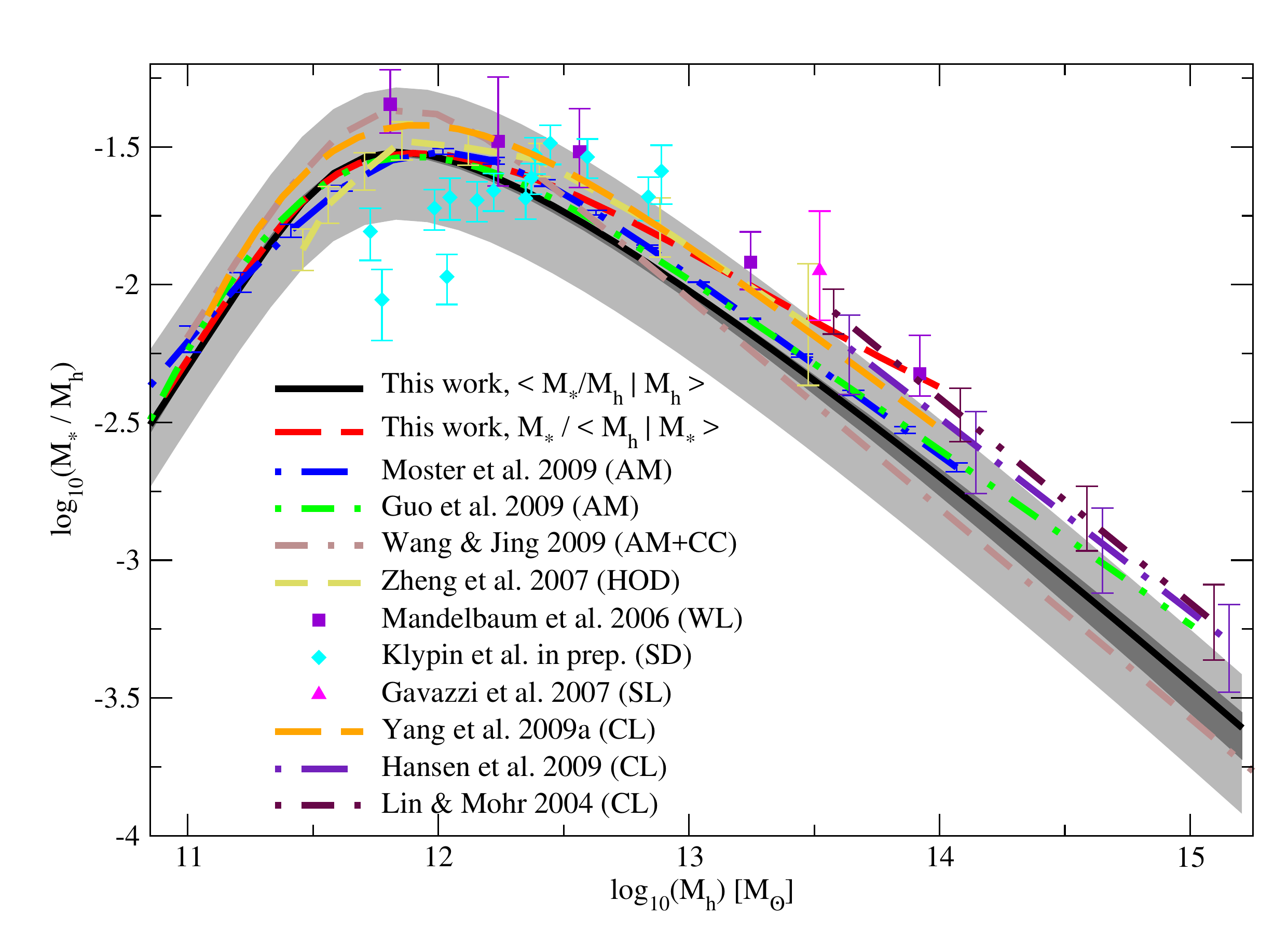}

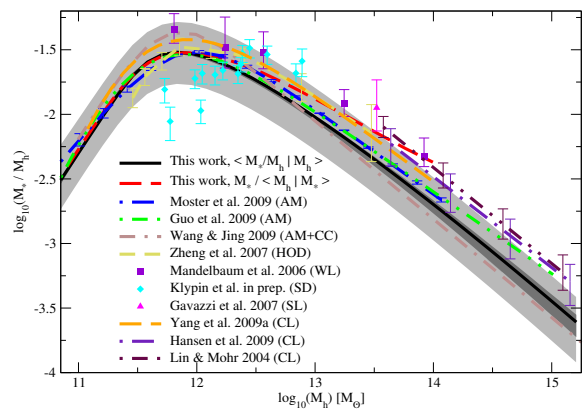
\captionof{figure}{Correlation between the stellar-to-halo mass ratio and halo
  mass. Even at the peak of the curve ($\sim$3\%) the stellar mass
  is significantly lower that the cosmic baryon to dark matter ratio,
  revealing an inefficient process of star formation. 
  Furthermore, the decrease of this fraction towards both the high-
  and low-mass end reveals the complexity of feedback mechanisms
  \citep[from][]{Behroozi:10}.}
\label{fig:AM}
\vskip+0.3truecm
\endgroup

A comparison of the observed stellar mass function of galaxies and
N-body simulations of dark matter halos (see Fig.~\ref{fig:AM})
suggests at least two distinct mechanisms to expel gas from galaxies,
one dominant at the low-mass end, and the other one controlling the
high-mass end. Since the efficiency of stellar winds is expected to
increase in weaker gravitational potentials, one would assume stellar
feedback is responsible for the low-mass trend. Similarly, the increasing
efficiency of AGN feedback with black hole mass would produce 
 the trend at the high-mass end. Furthermore, the strong
correlation between bulge mass (or velocity dispersion) and the mass
of the central supermassive black hole
\citep[Fig.~\ref{fig:MBHvSig}, see, e.g.][]{KorHo:13,Saglia:16} gives
further support to the role of AGN activity in shaping galaxy
formation.  However, this picture is too simplistic,

requiring a
better understanding of the physics.
Detailed analyses of winds driven by nearby starbursting galaxies
present a complex scenario that is not properly described by the
latest numerical codes of galaxy formation \citep{HB:15}. The
prevalence of outflows increases towards the younger phases of galaxy
formation. Therefore, detailed studies over complete samples during
the critical phases of galaxy evolution are needed to understand 
feedback in detail.

\begingroup
\centering
\includegraphics[width=85mm]{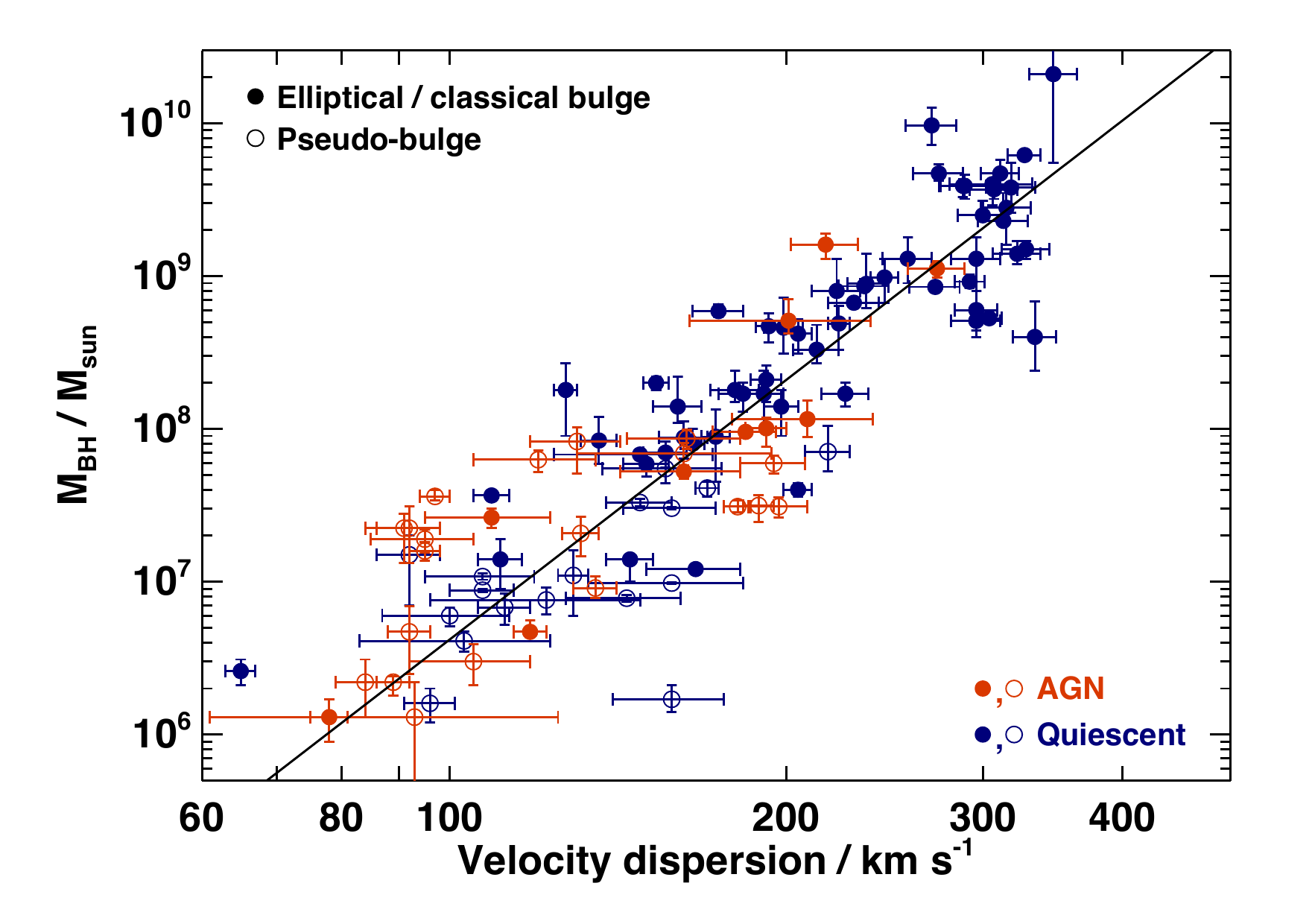}
\captionof{figure}{Correlation between black hole mass and velocity
  dispersion in local galaxies, from {\sl direct} measurements of the
  SMBH mass \citep[from][]{HB:14}.}
\label{fig:MBHvSig}
\vskip+0.2truecm
\endgroup

\begin{figure*}
\centering
\includegraphics[width=120mm]{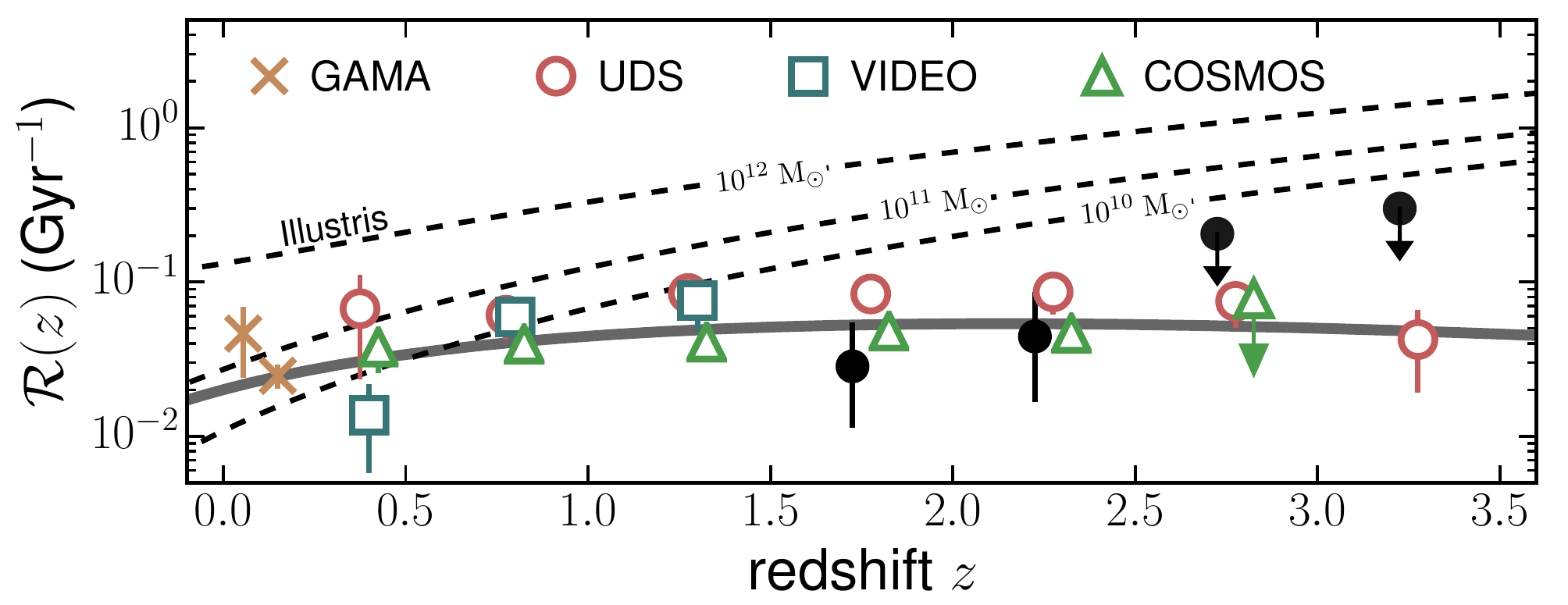}
\captionof{figure}{Redshift evolution of the merger rate, ${\cal R}(z)$,
  as measured by \citet{Mundy:17}. The
  observational constraints, shown as points with different
  symbols, are in stark contrast with respect to state-of-the-art
  predictions from the Illustris numerical simulations of galaxy
  formation \citep[dashed lines, labelled at different stellar masses;][]{Illustris}.}
\label{fig:Merg}
\vskip-0.3truecm
\end{figure*}

\subsection{Galaxy growth through mergers}

One of the main methods by which galaxies form is through the merger
process, whereby separate galaxies combine together to form a new
system. Merging is a significant channel of galaxy formation, and
needs to be measured with high precision if we are to understand how
galaxy formation proceeds. Closer to home, the complex structure of the
stellar populations found around the Milky Way, its vicinity and the
nearby Andromeda galaxy reflects the contribution of mergers to galaxy
growth \citep[e.g.][]{Ferguson:02,Ivezic:12}. Whilst mergers are
arguably not the way in which galaxies obtain the majority of their mass, this
process is still likely the main one for triggering AGN and black hole
formation and accounts for 25-50\% of the formation of massive galaxies
since z=3 \citep{Owns:14}.  Thus, a detailed quantitative assessment
of galaxy merger rates is a critical step that has not yet been fully
carried out, due to the lack of complete spectroscopic samples.
Furthermore, there are inconsistencies with the results
obtained so far and a disagreement with theory, showing that more
work, and better data are needed in this area.

Firstly, the exact role of mergers in galaxy formation is not clear,
with conflicting results, particularly at higher redshifts (z$>$1).  The
merger fraction at z$\simgt$1 is likely high, with a merger rate of
$\sim$0.5--1 mergers Gyr$^{-1}$ \citep[e.g.][]{Bluck:12,Tasca:15}.
Many merger rates at high redshift z$>$1 are measured with galaxy
structures, or based on samples of galaxies in kinematic or photometric redshift
pairs.  However, our best estimates of the merger rate differ from
theory by up to an order of magnitude (see Fig.~\ref{fig:Merg}
contrasting observational results with the latest, state-of-the-art
simulations by the Illustris collaboration). Moreover, we do not
have robust estimates about the role of minor mergers
in galaxy formation -- recovering
these will require very deep spectroscopic observations.

The best way to measure the merger rate at high redshift is through
spectroscopic pairs which requires both position and accurate radial
velocity information \citep[e.g.][]{CLS:12}.  However, the most up to
date studies have only used 12 pairs at z$>$2 to measure this
important quantity \citep{Tasca:14} with a merger fraction with
rather large errors ($19.4^{+9}_{-6}$\%) due to small number
statistics.

A near infrared spectroscopic survey of distant galaxies at z$>$1 will
give us the information we need to address this issue in detail.  A
survey with a high completeness level over the z=1--3 range will give
us a surface density over 10 times higher than previous surveys at
1$<$z$<$3 such as DEEP2, VVDS, and UDSz. To address this type of
science, the survey strategy needs to incorporate the option of
including such targets in the mask layout (if the method is to proceed
with a reconfigurable focal plane, see~\S\S\ref{SS:ReconfFP}). Given
the density of targets at the redshifts of interest, the merger
fraction will be measured to an accuracy an order of magnitude better
than what is currently known at these redshifts.

This is necessary to
ultimately pin down the amount of mass assembled through merging, as
well as to determine the role of merging on the triggering and
quenching of star formation, and on central AGN activity.  For
reference, in the most massive systems with M$_* > 10^{10}$M$_\odot$
it will be possible to measure merger fraction ratios of up to 1:30
down to a stellar mass limit of M$_* =10^{9.5}$M$_\odot$, such that we
can study, for the first time, the role of minor mergers in these
processes.

\subsection{The role of cold accretion}
\label{SS:ColdAcc}

The evolution of the gaseous component -- and its subsequent
transformation into stars -- is arguably one of the most complicated
problems in extragalactic astrophysics.
Hydrodynamical processes driving the gas flows, and feedback from star
formation, AGN activity or dynamical evolution of the baryon-dominated
central regions of halos lead to a significant mismatch between
the mass assembly history of dark matter halos, and the star formation
histories of galaxies embedded in these halos. In fact,
Fig.~\ref{fig:AM} illustrates this mismatch.

One key observable of the difference between dark matter growth and
galaxy growth is the presence of massive galaxies at early 
times \citep[e.g.][]{CI:04,McCarthy:04,Fontana:06,PPG:08}. A naive
mapping of dark matter growth into stellar mass growth leads 
to late star formation in massive galaxies, as found in
the first, pioneering computer simulations of galaxy formation
\citep[e.g.,][]{Kauff:96}. The presence of massive galaxies (stellar
mass $\simgt 10^{11}$M$_\odot$) with quiescent populations at
redshifts z$\sim$2-3 \citep[e.g.][]{FW4871} requires a mechanism by
which the commonly adopted process of star formation through
shock-heating of gas after the virialization of the halo, followed by
cooling \citep{RO:77} cannot be the main growth channel in these
systems.

\begingroup
\centering
\includegraphics[width=80mm]{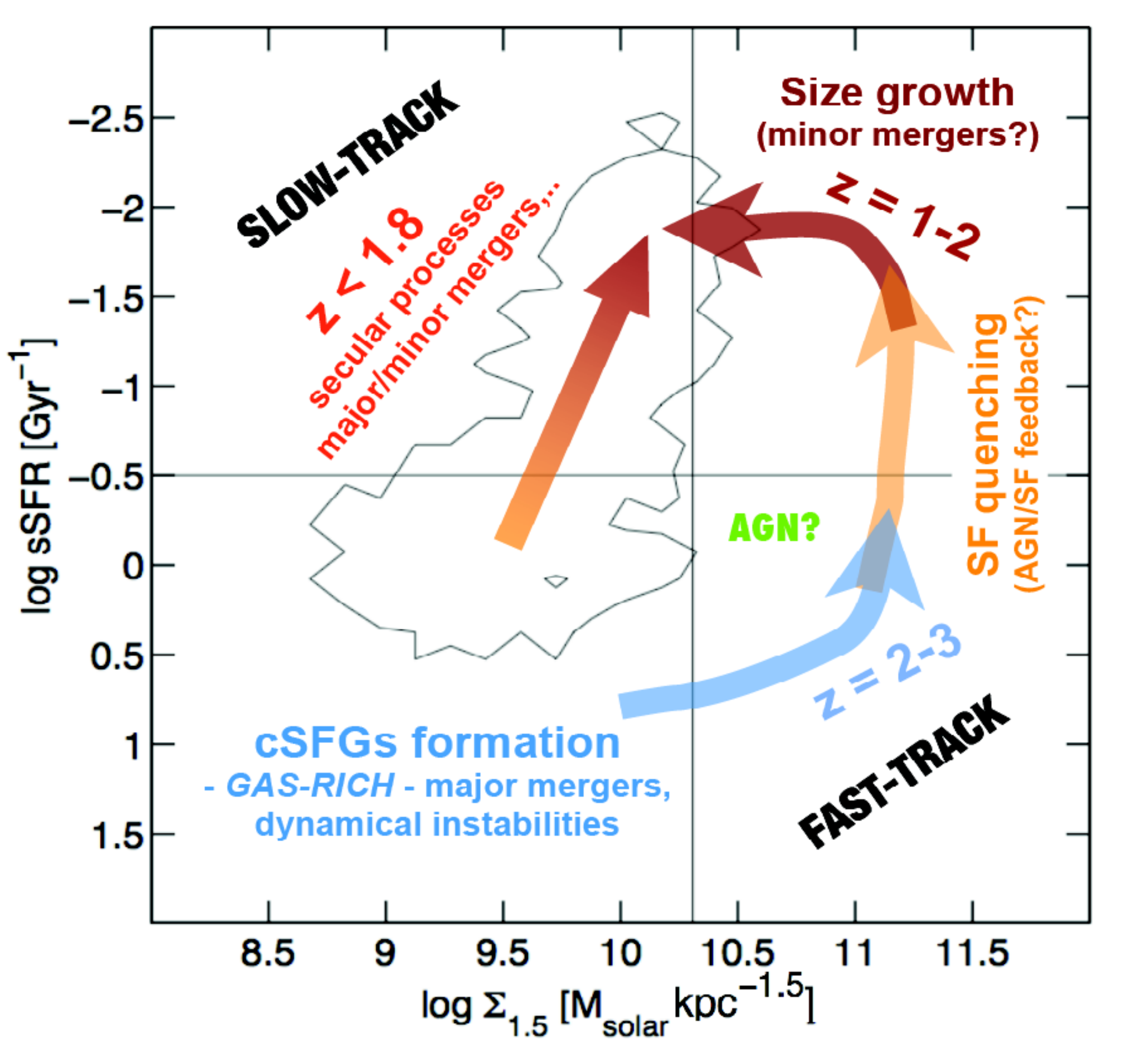}
\captionof{figure}{Mechanism proposed by \citet{Barro:13}
to explain the size evolution of massive galaxies.
The grey contour shows the galaxy distribution at low redshift. Two main
growth channels are proposed, involving a mixture of processes such as
merging, star formation quenching or secular processes. Large,
high quality spectroscopic data at these redshifts will allow us to
test in detail these proposals.}
\label{fig:SizeEv}
\vskip+0.2truecm
\endgroup

We find ourselves in a similar quandary with strong AGN
activity at very high redshift, z$\simgt$6 -- reflecting the presence
of very massive black holes within the first billion years of cosmic
time \citep{Fan:06}. In addition to the traditional hot-mode growth,
cold gas can also flow towards the centres of halos, following the
filamentary structure of the dark matter distribution, efficiently
feeding the central sites of star formation at early times
\citep{Dekel:09}.  Moreover, this process is found to operate in the
most massive systems at early times \citep{DB:06}.  Numerical
simulations suggest that clump migration and angular momentum transfer
provides an additional mechanism leading to the creation of massive
stellar cores at early times \citep{Ceve:10}.

However, observational constraints of the role of cold accretion are
few, and no conclusive evidence has been found to date. A large
spectroscopic galaxy survey probing the peak of evolution would allow
us to study the hot- and cold-mode growth channels of star formation
and black hole growth, and the connection with redshift and
environment. A detailed analysis of the shape of targeted spectral
lines will allow us to detect and quantify gas inflows, but a large
volume of data is necessary given the small covering factor
of accretion flows \citep{FG:11}. The high S/N
of this survey will make studies of individual galaxies (not stacked
spectra) available. As of today, state-of-the-art samples comprise
$\sim$100 spectra with just enough S/N to study bright emission lines
(see, e.g. \citealt{Genzel:14} with VLT/KMOS; or \citealt{Kacprzak:16}
with Keck/MOSFIRE). These studies give promising results about the
presence of this important process of galaxy growth. Note studies
in the Ly-$\alpha$ region (i.e. concerning the {\sl cosmic dawn} survey,
\S\ref{sec:Lyman}) can also be used to obtain constraints
on gas inflows \citep{Yajima:15}.

\subsection{Size evolution}

An additional conundrum raised by the study of massive galaxies at
high redshift is the issue of size evolution. The comoving number
density of massive ($\simgt 10^{11}$M$_\odot$) galaxies has been found
not to decrease very strongly with redshift (z$\simlt 2$), with respect
to the predictions from simple models of galaxy formation that mostly
link galaxies to the evolution of the dark matter halos
\citep[e.g.][]{CC:07,IF:09}.

This would reflect an early formation of these type of galaxies,
whereby the bulk of the stellar mass is in place by redshift
z$\sim$2--3. However, the sizes of these galaxies at z$\simgt$1--2 are
significantly smaller than their low-redshift counterparts
\citep[e.g.][]{Daddi:05,Trujillo:06}. A large volume of publications
has been devoted to propose mechanisms that could explain this puzzle,
including gas outflows as a mechanism to alter the gravitational
potential, ``puffing-up'' the dense central region.

However, the (old) stellar populations typically found in massive
galaxies do not allow for significant quantities of recent star
formation, or cold gas flows to explain this size evolution
\citep{I3}, suggesting instead a growth process through gas-free (dry)
merging. This merging can proceed dramatically -- through a small
number of major mergers \citep{KS:09}, where the merging progenitors
have similar mass -- or through a more extended and smooth process of
minor merging \citep{Naab:09}. In addition, one should consider
whether these evolved compact cores end up as massive (and extended)
early-type galaxies in high density regions \citep{Pogg:13}, or as
massive bulges of disk galaxies \citep{IGDR:16}.

Fig.~\ref{fig:SizeEv} shows a diagram of how this may work, from an analysis of
massive galaxies in CANDELS \citep{Barro:13}, with an interesting
evolution from massive compact systems with a strong star
formation rate, towards the quiescent galaxies we see today, involving
both secular processes, galaxy mergers and star formation
quenching. Establishing such connections requires a large volume
of galaxy spectra at the peak of galaxy formation activity.
All these studies are based on relatively small samples
($\simlt 10^3$) with mostly high-quality photometry (from HST) but comparatively poor
spectroscopic data. {\sl Accurate characterization of the stellar
population content of these galaxies will enable us to robustly
constrain the processes by which galaxies grow.}

\subsection{Reaching out: the role of environment}
The environment where galaxies reside plays a significant
role in shaping their observed properties and thus their evolution. It
essentially deprives them of their hot and cold gas reservoirs, thus
quenching their star formation activity, and also can literally
disrupt them by removing their stars \citep{AP:15}.  The observed
properties of galaxies in the local Universe have provided us with a
wealth of evidence towards environmental processes, whose time scales
and amplitudes are unfortunately known only at a qualitative level. A
robust quantitative estimate of the dependence of such parameters on
environment and redshift largely remains an open problem.

The Sloan Digital Sky Survey (SDSS) has been the very first survey to
perform an unprecedented and statistically significant census of the
photometric and spectroscopic properties of z$\simeq$0 galaxies at optical
wavelengths. It has permitted us to detail star formation activity in
galaxies across several orders of magnitude, with respect to galaxy
stellar mass, environment and infall time at z$\simeq$0. We are now aware that the
number of quenched galaxies -- not forming new stars any longer --
rises with their stellar mass at a fixed kind of environment, and with
environment magnitude (from small galaxy groups to large clusters) at
fixed stellar mass \citep{Wein:06,VdB:08,AP:09,Wetz:12}.

\begingroup
\vskip+0.2truecm
\centering
\includegraphics[width=75mm]{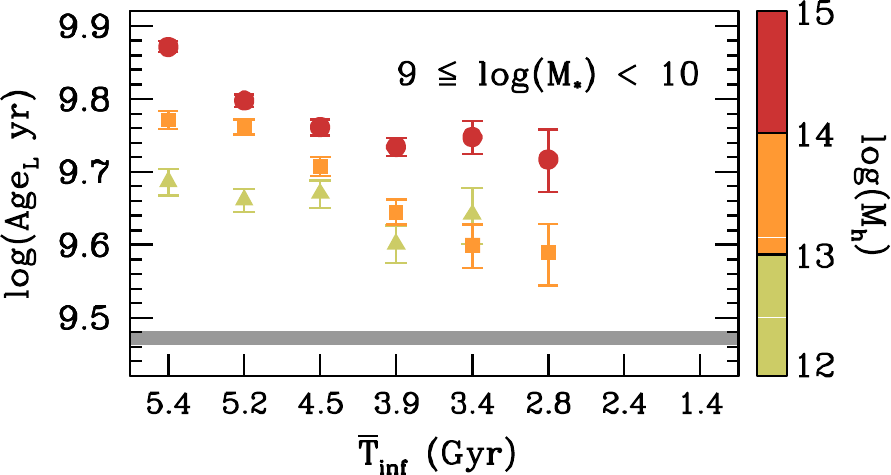}
\captionof{figure}{The stellar age of galaxies less massive
  than $10^{10}$M$_\odot h^{-2}$ is shown as a function of infall
  time. Galaxies are colour-coded regarding halo mass: red circles,
  orange squares and yellow triangles identify galaxies in clusters,
  rich groups and low-mass groups, respectively. The grey stripe
  indicates the  stellar age of equally-massive galaxies in the
  field \citep[from][]{AP:19}.}
\label{fig:Env}
\vskip+0.2truecm
\endgroup

We have also learnt from SDSS that the age of the bulk of stars in a
galaxy grows progressively older i) as their infall time increases
(i.e. galaxies accreted onto their present-day host environment early
on are now older than those accreted more recently); 2) as their
environment, at fixed infall time, becomes more massive, indicating
that the environment mass enhances the efficiency of those physical
processes able to quench star formation in galaxies
\citep{AP:10, AP:19, Smith:19}.  In addition, recently accreted cluster galaxies
appear to be older than equally-massive field galaxies, an
observational result that has been attributed to group-preprocessing:
the star-formation quenching of these recent infallers started already
while they were still living in smaller groups, that later merged with
clusters \citep[see Fig.~\ref{fig:Env},][]{AP:19}. Such group-preprocessing has extensively been
advocated by semi-analytic models of galaxy formation and evolution in
order to explain the large number of quenched galaxies observed in
clusters \citep{DeLuc:12,Wetz:13}

The observational evidence described above highlights the importance
of knowing the accretion epoch of a galaxy if we want to understand
the role of environment. Unfortunately, we can not use observations of
z$\simeq$0 galaxies to accurately derive their infall epochs when they
became exposed to environmental effects for the first time.  To
determine such an important moment in the evolution of galaxies we
need to quantify and study environment at different redshifts; this is
what a deep-wide NIR spectroscopic galaxy survey will enable us to do,
by tracing the assembly history of environments with cosmic time,
providing us with a direct measurement of the redshift of infall of
galaxies as a function of their stellar mass.  Moreover, the lensing
and X-ray information from {\sl Euclid} and {\sl eRosita},
respectively, combined with the accurate spectroscopic information
produced by {\tt Chronos} will probe the dependence of the star
formation histories on the dark matter halos. While the data from {\sl
  Euclid} and {\sl eRosita} will mainly target the assembly of massive
environments, thus introducing a significant bias towards star-forming
galaxies, {\tt Chronos} will broaden the study to smaller environments
and consequently will avoid the selection bias of the {\sl Euclid}
sample.

When and in which environments did the quenching of the star formation
activity of galaxies start? How fast did it proceed? The quantitative
and direct replies to these inquiries are provided by our measurements
of star formation rates, star formation histories and chemical
enrichment of galaxies of different stellar mass, in different
environments at different epochs, from z$\sim$1--3 to z=0. Only these
observables allow us to directly estimate the typical time scales of
star formation in galaxies, and to achieve a model-independent value
of the time scales over which galaxy groups and clusters switched
galaxy star formation off, and produced the observed present-day
galaxy populations.

With increasing redshift these measurements shift to infrared
wavelengths and become challenging even for modern ground-based
telescopes. Ground-based measurements allow for only a partial
characterization of the properties of galaxies at z$>$0.5, for which we can
mostly measure emission lines (thus star formation rates) since their
absorption lines (used as age and metallicity indicators) become less
and less accessible. The data gathered so far on galaxies at $0.3<$z$<0.8$
indicate that the fraction of quenched galaxies is larger in
galaxy groups than in the field, but definitively lower than the
fraction of quenched galaxies in groups at z$\simeq$0 \citep{Wil:05,McGee:11}.
At intermediate redshifts, the fraction of star forming galaxies
diminishes from 70-100\% in the field to 20-10\% in the more massive
galaxy clusters \citep{Pogg:06}. However, the star formation rates of
group galaxies do not significantly differ from those in the field;
only star forming galaxies in clusters show star formation rates a
factor of 2 lower than in the field at fixed stellar mass
\citep{Pogg:06,Vul:10,McGee:11}.

At the highest redshifts probed for environment,
$0.8<$z$<1$, the more massive galaxy groups and clusters are mostly populated
by quenched galaxies and both exhibit a 30\% fraction of
post-starburst galaxies \citep[i.e. with a recently truncated star
  formation activity,][]{Bal:11}.  In particular, the fraction of
post-starburst galaxies in clusters exceeds that in the field by a
factor of 3. Cluster and field galaxies still able to form new stars
share instead similar star formation rates.  On the basis of these
results, \citet{Muzz:12} have argued that, at z$\sim$1, either
the quenching of star formation due to the secular evolution of
galaxies is faster and more efficient than the quenching induced by
galaxy environment, or both mechanisms occur together with the same
time scale. Which mechanism prevails and over which time scale?
At present, we do not know.

To further progress on this issue, we require a facility such as
{\tt Chronos} to observe a complete stellar-mass limited sample of
environments at z$\geq$1--3, and to derive the star formation histories
of their galaxies with an unprecedented accuracy. {\tt Chronos}
observations will thus deliver the fading time scales of star
formation of galaxies of different stellar mass residing in groups and
clusters.  This is not simply an incremental step in our knowledge of
environment-driven galaxy evolution.  This is the {\sl still missing,
fundamental quantitative change} from the simple head-count of
quenched or star-forming galaxies to the measurement of physical
properties of galaxies in environments at {\sl cosmic noon}.
\end{multicols}

\section{First galaxies and the epoch of reionization}
\label{sec:Lyman}
\begin{multicols}{2} 

\subsection{Leaving the dark ages}

Cosmic reionization is a landmark event in the history of the
Universe. It marks the end of the ``Dark Ages'', when the first stars
and galaxies formed, and when the intergalactic gas was heated to tens
of thousands of Kelvin from much colder temperatures.  This
global transition, during the first billion years of cosmic history,
had far-reaching effects on the formation of early cosmological
structures and left deep impressions on subsequent galaxy and star
formation, some of which persist to the present day.

The study of this epoch is thus a key frontier in completing our
understanding of cosmic history, and is currently at the forefront of
astrophysical research \citep[e.g.][]{Robertson:15}. Nevertheless,
despite the considerable recent progress in both observations and
theory (e.g. see recent reviews by \citealt{Dunlop:13} and \citealt{Loeb:13})
all that is really established about this crucial era is that Hydrogen
reionization was completed by redshift z$\sim$6 (as evidenced by
high-redshift quasar spectra; \citealt{Fan:06}) and probably commenced
around z$\sim$12 (as suggested by the {\sl Planck} 
polarisation measurements, which favour a `mean' redshift of
z$_{\rm re} = 8.8^{+1.7}_{-1.4}$\,; \citealt{Planck:15}).  However, within
these bounds the reionization history is essentially unknown. New
data are required to construct a consistent picture of reionization
and early galaxy formation/growth (see Fig.~\ref{fig:reioniz}).

Understanding reionization is therefore a key science goal for a
number of current and near-future large observational projects. In
particular, it is a key science driver for the new generation of major
low-frequency radio projects (e.g. LOFAR, MWA and SKA) which aim to
map out the cosmic evolution of the neutral atomic Hydrogen via 21-cm
emission and absorption. However, such radio surveys cannot tell us
about the sources of the ionizing flux, and in any case radio
observations at these high redshifts are overwhelmingly difficult, due
to the faintness of the emission and the very strong foregrounds.  It
is thus essential that radio surveys of the neutral gas are
complemented by near-infrared surveys which can both map out the
growth of ionized regions, and provide a complete census of the
ionizing sources.

\begingroup
\vskip+0.2truecm
\centering
\includegraphics[width=75mm]{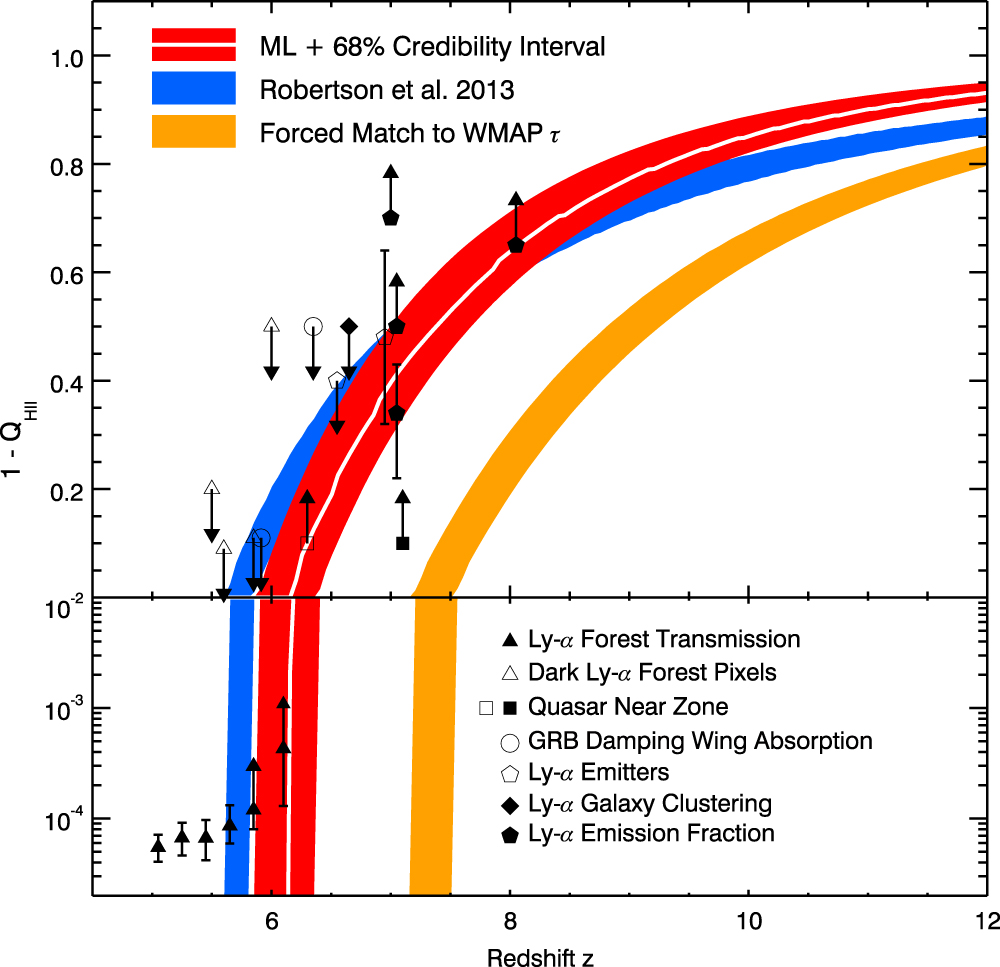}
\captionof{figure}{Measures of the neutrality
  $1-{{Q}_{{{{\rm H}}_{{\rm II}}}}}$ of the intergalactic medium as a
  function of redshift. Shown are the observational constraints, along
  with model predictions of the evolving IGM neutral fraction (in
  red). The bottom panel shows the IGM neutral fraction near the end
  of the reionization epoch, where the model fails to capture the
  complexity of the reionization process.
  \citep[from][]{Robertson:15}.  }
\label{fig:reioniz}
\vskip+0.2truecm
\endgroup

A genuine multi-wavelength approach is required,
and cross-correlations between different types of observations will be
necessary both to ascertain that the detected signals are genuine
signatures of reionization, and to obtain a more complete
understanding of the reionization process.
It has thus become increasingly clear that a wide-area, sensitive,
spectroscopic near-infrared survey of the z=6--12 Universe is required
to obtain a proper understanding of the reionization process and early
galaxy and black-hole formation.  Such a survey cannot be undertaken
from the ground (due to Earth's atmosphere), nor with {\sl JWST} (inadequate
field-of-view), nor {\sl Euclid} or {\sl WFIRST} (inadequate
sensitivity with slitless spectra). Only a mission such as
{\tt Chronos} can undertake such a survey and simultaneously address the
three, key, interelated science goals which we summarize
below. Moreover, detailed studies of z$>$6 galaxies in the Ly-$\alpha$
region will complement the information provided at longer wavelengths
by ALMA \citep[e.g.][]{Capak:15}.

\subsection{The clustering of Ly-$\alpha$ emitters as a probe of reionization}

Cosmological simulations of reionization predict that the
highly-clustered, high-redshift sources of Lyman-continuum photons
will lead to an inhomogeneous distribution of ionized regions.  The
reionization process is expected to proceed inside-out, starting from
the high-density peaks where the galaxies form. Thus, as demonstrated
by the state-of-the-art simulations shown in Fig.~\ref{fig:reioniz2},
reionization is predicted to be highly patchy in nature.  This
prediction is already gaining observational support from the latest
large-area surveys for Ly-$\alpha$ emitters at z$\sim$6.5, where it has been
found that, depending on luminosity, their number density varies by a
factor of 2--10 between different $\frac{1}{4}$\,deg$^2$ fields
\citep{Ouchi:10,Nakamura:11}. It is thus clear that surveys over many
square degrees are required to gain a representative view of the
Universe at z$>$6. Crucially, with such a survey, the differential
evolution and clustering of Lyman-break galaxies and Ly-$\alpha$
emitting galaxies can be properly measured for the first time,
offering a key signature of the reionization process.

High-redshift galaxies can be selected on the basis of
either their redshifted Lyman break (the sudden drop in emission from
an otherwise blue galaxy, due to inter-galactic absorption at
wavelengths $\lambda_{\rm rest} < 1216$\AA), or their
redshifted Ly-$\alpha$
emission. The former class of objects are termed Lyman-Break Galaxies
(LBGs) while the latter are termed Ly-$\alpha$ Emitters (LAEs).
In principle, LAEs are simply the subset of LBGs with detectable
Ly-$\alpha$ emission, but the current sensitivity limitations of
broad-band near-infrared imaging over large areas has meant that
narrow-band imaging has been successfully used to yield samples of
lower-mass galaxies which are not usually identified as LBGs
\citep[e.g.][]{Ono:10}.  Nevertheless, as demonstrated by
spectroscopic follow-up of complete samples of bright LBGs
\citep[e.g.][]{Stark:10, Vanzella:11,Schenker:12}, the fraction of
LBGs which are LAEs as a function of redshift, mass, and environment
is a potentially very powerful diagnostic of both the nature of the
first galaxies, and the physical process of reionization.

\begingroup
\vskip+0.2truecm
\centering
\includegraphics[width=75mm]{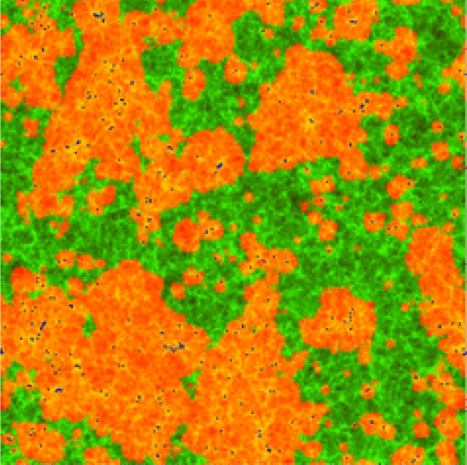}
\captionof{figure}{The geometry of the epoch of reionization, as
  illustrated by a slice through a (165\,Mpc)$^3$ simulation volume at
  z=9. Shown are the density (green/yellow), ionized fraction
  (red/orange), and ionizing sources (dark dots) \citep{Iliev:12}.
  The necessity of a deep, near-infrared spectroscopic survey
  covering many square degrees is clear.}
\label{fig:reioniz2}
\vskip+0.2truecm
\endgroup

With the unique combination of deep, wide-area near-infrared imaging,
provided by surveys such as {\sl Euclid} and {\sl WFIRST}, and deep, complete
follow-up near-infrared spectroscopy, made possible with {\tt Chronos}, we
propose to fully exploit the enormous potential of this
approach.

The essential idea of using {\tt Chronos} to constrain
reionization is as follows: while the Ly-$\alpha$ luminosity of LAEs is
affected both by the intrinsic galaxy properties, and by the \HI
content (and hence reionization), the luminosity of LBGs (which is
measured in the continuum) depends only on the intrinsic galaxy
properties. Thus, a deep, wide-area, complete survey for LBGs at
z$\sim$6--12 with accurate redshifts secured by {\tt Chronos} will deliver a
definitive measurement of the evolving luminosity function and
clustering of the emerging young galaxy population, while the analysis
of the follow-up spectroscopy will enable us to determine which LBGs
reside in sufficiently large ionized bubbles for them to also be
observed as LAEs. In order to prevent strong damping wing absorption
of Ly-$\alpha$ photons, a galaxy must carve out a bubble of radius R$_I$ of
500--1000 physical kpc at z$\sim$8. According to the most recent
reionization history predictions from cosmological simulations,
consistent with the various reionization constraints, the \HI fraction
at this redshift is around $\chi \sim 0.5$--$0.7$. R$_I$ for a typical
galaxy with a star-formation rate of $\dot{\rm M}_* = 1$\,M$_\odot$\,yr$^{-1}$
is expected to be considerably smaller (though it depends on poorly
established values of the ionizing photon escape fraction; cf.
\citealt{RM:03}). Thus, such galaxies will be only marginally
detectable in the Ly-$\alpha$ line if they are isolated. In practice, some of
these galaxies will be highly clustered and therefore will help each
other in building a \HII region which is large enough to clear the
surrounding \HI and make it transparent to Ly-$\alpha$ photons.

This argument emphasizes the importance of clustering studies of LAEs,
for which the proposed survey is optimally designed.  A key aim is to
compute in great detail the two-point correlation function of LAEs and
its redshift evolution. For the reasons outlined above, reionization
is expected to increase the measured clustering of emitters and the
angular features of the enhancement would be essentially impossible to
attribute to anything other than reionization. 

In fact, under some scenarios, the apparent clustering of LAEs can be
well in excess of the intrinsic clustering of halos in the concordance
cosmology. Observing such enhanced clustering would confirm the
prediction that the \HII regions during reionization are large
\citep{McQuinn:07}.  As required to meet our primary science goals,
the {\tt Chronos} surveys will result in by far the largest and most
representative catalogues of LBGs and LAEs ever assembled at
z$>$6. Detailed predictions for the number of LBGs as extrapolated
from existing ground-based and HST imaging surveys are deferred to the
next subsection.  However, here we note that the line sensitivity of
the 100\,deg$^2$ spectroscopic survey will enable the identification
of LAEs with a Ly-$\alpha$ luminosity $\ge 10^{42.4}$\,erg\,s$^{-1}$,
while over the smaller ultra-deep 10\,deg$^2$ survey this
line-luminosity limit will extend to $\ge 10^{41.6}$\,erg\,s$^{-1}$.
Crucially this will extend the Ly-$\alpha$ detectability of LBG
galaxies at z$\sim$8, with brightness J$\sim$27AB, down to ``typical''
equivalent widths of $\sim$15\AA\ \citep{Stark:10,Vanzella:11,CL:12,
  Schenker:12}.

The total number of LAEs in the combined surveys (100 + 10\,deg$^2$)
will obviously depend on some of the key unknowns that {\tt Chronos} is
designed to measure, in particular the fraction of LBGs which display
detectable Ly-$\alpha$ emission as a function of redshift, mass and
environment. However, if the observed LAE fraction of bright LBGs at
z$\sim$7 is taken as a guide, the proposed surveys will uncover $\sim$10,000
LAEs at z$>$6.5.

\subsection{The emerging galaxy population at z$>$7, and the supply of reionizing photons}

The proposed survey will provide a detailed spectroscopic
characterization of an unprecedently large sample of LBGs and
LAEs. Crucially, as well as being assembled over representative
cosmological volumes of the Universe at z$\sim$6--12, these samples
will provide excellent sampling of the brighter end of the galaxy UV
luminosity function at early epochs. As demonstrated by the most
recent work on the galaxy luminosity function at z$\sim$7--9
\citep{McLure:13}, an accurate determination of the faint-end slope of
the luminosity function (crucial for understanding reionization) is in
fact currently limited by uncertainty in L$_*$ and
$\Phi_*$. Consequently, a large, robust, spectroscopically-confirmed
sample of brighter LBGs over this crucial epoch is required to yield
definitive measurements of the evolving luminosity functions of LBGs
and LAEs.

Leaving aside the uncertainties in the numbers of LAEs discussed
above, we can establish a reasonable expectation of the number of
photometrically-selected LBGs which will be available within the
timescales expected for such a mission. For example, scaling from
existing HST and ground-based studies, the ``Deep'' component of the
{\sl Euclid} survey (reaching J$\sim$26AB at 5$\sigma$ over $\sim$40\,deg$^2$),
is expected to yield $\sim$6000\,LBGs in the redshift range 6.5$<$z$<$7.5
with J$<$26AB (selected as ``z-drops''), $\sim$1200 at 7.5$<$z$<$8.5
(``Y-drops''), and several hundred at z$>$8.5 (``J-drops'')
\citep{Bouwens:10,Bowler:12,McLure:13}.

Therefore, the planned spectroscopic follow-up over 10\,deg$^2$, will be
able to target (at least) $\sim$1500\,LBGs in the redshift range 6.5$<$z$<$7.5,
$\sim$300 in the redshift bin 7.5$<$z$<$8.5, and an as yet to be
determined number of candidate LBGs at 8.5$<$z$<$9.5. The proposed
depth and density of the {\tt Chronos} near-infrared spectroscopy will allow
detection of Ly-$\alpha$ line emission from these galaxies down to a 5$\sigma$
flux limit $10^{-18}$\,erg\,cm$^{-2}$\,s$^{-1}$, enabling rejection of any low-redshift
interlopers, determination of the LAE fraction down to equivalent widths of $\sim$10\AA,
and accurate spectroscopic redshifts for the LAE subset.

\subsection{The contribution of AGN to reionization \& the early growth of black holes}

SDSS has revolutionised studies of quasars at the highest redshifts,
and provided the first evidence that the epoch of reionization was
coming to an end around z$>$6 \citep{Becker:01}.  As with the
studies of galaxies discussed above, pushing to higher redshifts is
impossible with optical surveys, regardless of depth, due to the fact
that the Gunn-Peterson trough occupies all optical bands at z$>$6.5.
Therefore, to push these studies further in redshift needs deep
wide-field surveys in the near-infrared.

The wide-area, ground-based VISTA near-infrared public surveys such as
VIKING and the VISTA hemisphere survey are slowly beginning to uncover
a few bright quasars at z$\sim$7 \citep[e.g.][]{Mortlock:11}.  Recent
evidence combining X-ray and near-IR data suggests that faint quasars
at z$\sim$6 may be commoner than previously thought, and might contribute
to reionization significantly \citep{Giallongo:15,MH:15}.
It is expected that {\sl Euclid} and {\sl WFIRST} will be able to provide
a good determination of the bright end of the QSO luminosity function
at z$>$6. However, the shape of the QSO luminosity function at these
redshifts can only be studied with detailed near-infrared spectroscopy
over a significant survey area. This is the only direct way to
properly determine the contribution of accreting black holes to the
reionization of the Universe and constrain the density of black-holes
within the first Gyr after the Big Bang; the combination of depth and
area proposed in this NIR survey provides the ideal way in which to
measure the evolving luminosity function of quasars at 6.5$<$z$<$10.
\end{multicols}

\eject

\section{Precision Cosmology}
\label{sec:cosmo}

\begin{multicols}{2} 

The {\sl Euclid} mission will revolutionize cosmology, however the ultimate
precision of {\sl Euclid} will be limited by our understanding of galaxy
evolution on small-scales ($\simlt$1\,Mpc) due to baryonic feedback
mechanisms. For example \citet{vDaal:11} predicted that
AGN feedback should have a sizeable 20\% effect on the amplitude of
the matter power spectrum, amongst many other studies. Without
calibration data on small-scales from large complete spectroscopic
samples, {\sl Euclid} will be required to either marginalize over such
effects, remove them from the analyses using filter techniques, or
model them using a phenomenological ansatz such as the halo model.

Understanding galaxy evolution will therefore enable precision
cosmology to be extended beyond the {\sl Euclid} baseline to smaller scales, 
allowing for an increased sensitivity of modified
gravity models, and up to a ten fold improvement on dark energy
constraints than from {\sl Euclid} alone. As example of beyond-{\sl Euclid}
cosmology enabled by small-scale information, we list the following:

\begin{itemize}
\item Neutrino Physics. Massive neutrinos impact the matter
  power spectrum on both linear and non-linear scales. In particular, 
  information on the neutrino hierarchy is amplified on small-scales
  \citep{Jimenez:10}.

\item Warm Dark Matter. The temperature, and particle mass,
  of dark matter is still unknown. In fact models in which dark matter
  has a small temperature are still allowed by the data. If dark
  matter is warm then any signature of its effects will be seen on
  small-scales, e.g. in the stellar mass function.

\item Modified Gravity. The accelerated expansion could be
  a symptom of our gravity model, general relativity, being
  incorrect. Models that change general relativity can have a
  scale-dependence, and chameleon mechanisms can act on relatively
  small scales \citep{Amen:13}.
\end{itemize}
    
Figure~\ref{fig:Cosmo} shows the sensitivity of three beyond-{\sl Euclid}
cosmological models to small-scale information. The deep redshift
range would also constrain early-dark energy models, complementing the
{\sl Euclid} cosmology objectives using techniques such as those used by
\citet{Mandel:12} in SDSS.

\begingroup
\centering
\includegraphics[width=80mm]{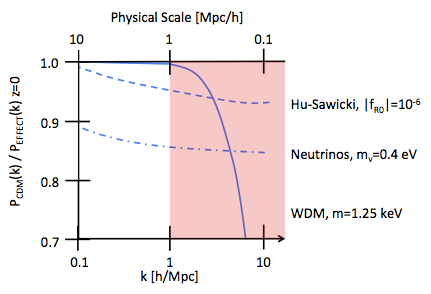}
\captionof{figure}{A sample of new physical effects that can be tested using
  small-scale information. Shown is the ratio of the unaffected power
  spectrum compared to that with the new physical effects, as a
  function of scale, at redshift zero. The solid line
  (from \citealt{Markovic:11}) shows the impact of a 1.25\,keV warm dark matter
  particle, The dot-dashed line shows the impact of a massive
  neutrinos with total mass of 0.4\,eV from \citet{Zhao:13}, and the
  dashed line shows a Hu-Sawicki modified gravity model with an
  amplitude deviation in the Lagrangian of $10^{-6}$ \citep{Baldi:14}.}
\label{fig:Cosmo}
\endgroup

\end{multicols}

\section{Scientific requirements}
\label{sec:scireq}
\begin{multicols}{2} 

\subsection{Introduction}
The study of galaxy formation and evolution involves a large range of
measurement concepts. A deep spectroscopic galaxy survey -- combined
with high resolution NIR imaging from 
{\sl Euclid} and {\sl WFIRST} -- provides the optimal dataset. Note, however, the
inherently more complex task of gathering high-quality
spectroscopic data with respect to imaging.  A spectral resolution
$R\equiv\lambda/\Delta\lambda\sim 1500-3000$ is needed both for 
accurate velocity dispersion measurements, 
and to beat the degeneracies present in spectral features. This limit is
mainly set by the typical stellar velocity dispersions found in
galaxies (50--300\,km\,s$^{-1}$), and by the need to adequately
resolve targeted emission lines and absorption features.  Fig.~\ref{fig:Hlimit} quantifies the
magnitude limit within the targeted redshift range.  Ideally, a H=26AB
limit, {\sl in the continuum}, would provide complete samples down to
a stellar mass of M$_*\simgt 10^9$M$_\odot$ across the peak of
galaxy formation activity (z$\sim$1--3). Note that at
higher redshifts, the analysis will rely on emission lines, although
it will be possible to work in the continuum of the most massive
galaxies (M$_*\simgt 10^{9.5}$M$_\odot$ at z$\sim6$).

Regarding the issue of target selection for spectroscopy, H=26AB is
the sensitivity limit expected for the deep fields with {\sl
  Euclid}/NISP \citep{Euclid}, and {\sl WFIRST}/WFI will provide
photometry slightly deeper than this \citep{WFIRST}. 

\begingroup
\vskip+0.3truecm
\centering
\includegraphics[width=85mm]{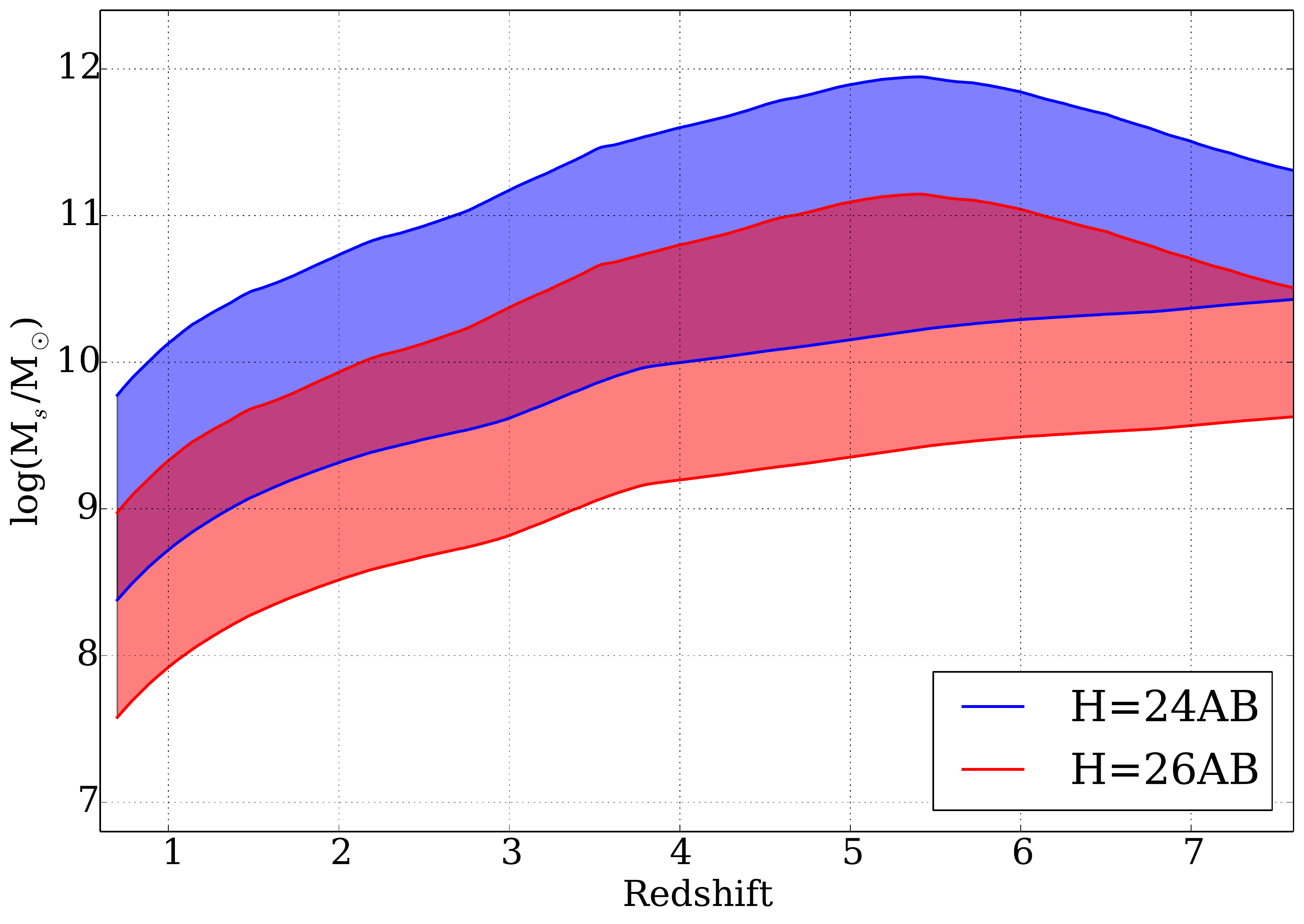}
\captionof{figure}{Stellar mass of a range of stellar populations with respect
  to redshift, with apparent magnitude H=24AB and H=26AB
  \citep[from the synthetic models of ][for a Chabrier IMF, at solar metallicity]{BC:03}.
  The shaded regions extend from old stellar populations (formed at $z_{\rm FOR}=10$) to
  a younger galaxy (age 50\,Myr). Real galaxies will mostly sit within the shaded
  regions. 
}
\label{fig:Hlimit}
\vskip+0.3truecm
\endgroup

Note the highly challenging measurements: at the faint 
end, a H=26AB distant galaxy produces a flux of $\sim$3
photons per second in a perfect, unobstructed 3\,m diameter telescope
through the WFC3/F160W passband. Furthermore, the same collecting area
yields $\sim$30 photons per {\sl hour}, per spectral resolution
element, in the continuum of a spectrum at R=2000. The
sky brightness at the best ground-based sites reach
$\mu_{\rm H,AB}^{\rm Sky}\sim$19.5\,mag\,arcsec$^{-2}$ \citep{SkyH}, and the
zodiacal background can be as high as $\mu_{\rm H,AB}^{\rm Zodi}\sim$21.5\,mag\,arcsec$^{-2}$
in the same spectral region\footnote{Wide Field Camera 3 Instrument Handbook
  for cycle 24 (STScI, v8.0, Jan 2016)}.
At these limiting magnitudes, any successful project must be 
based in space, and requires very long integration
times, pointing towards the darkest regions away from the galactic plane
and the ecliptic. For reference, the best spectroscopic samples of galaxies at
z$\sim$2--3 with state-of-the-art, ground-based facilities
(e.g. VLT/X-SHOOTER) reach K$\simlt$21.5{\sc AB}, and have noisy
continua \citep[e.g.][]{Marsan:16}.

In a presentation for the future
ESA L2/L3 science cases \citep{Chronos}, we argued that {\sl any}
ground-based facility, including future telescopes such as {\sl ELT}
or {\sl TMT}, will not be capable of providing a clean spectrum over
a wide spectral window, needed to trace in
detail the continuum associated to the stellar populations of galaxies
at the peak of formation.

\begin{center}
{\sl \vskip-2truemm Tentative mission concept}
\end{center}
\vskip-3truemm

The proposed science case will require a
large aperture survey telescope in space (between 3 and 6\,m
diameter), ideally at L2, although bolder options in the future may
consider a lunar platform (allowing for service missions,
and providing added value to a future manned programme to the moon).
The survey will entail 
long total integration times per field, over the 100\,ks mark -- requiring fine
pointing accuracy. {\sl Such a survey would be, by
far, the deepest ever taken.} The baseline concept proposed in
\citet{Chronos} was equivalent to taking one Hubble Ultra Deep Field
every fortnight for five years. 
Such characteristics places {\sl Chronos} as an L-type mission, ideally
including cross-collaborations with international space agencies outside
the ESA domain. A smaller, M-type, mission could be envisioned for technology
development, targetting the most luminous galaxies in the two cosmic
intervals under study.

\subsection{Why target one million spectra?}
\label{SS:WhySoMany}
The aim of the survey is to provide a legacy database of high quality
galaxy spectra, sampling both the peak (z$\sim$1--3) as well as the first
phases (z$\simgt$6) of galaxy formation. In contrast with
cosmology-driven missions -- such as {\sl Euclid} or {\sl Planck} -- 
that have a unique figure of merit for the constraint of a reduced set
of cosmological parameters, {\tt Chronos} will be a
``general-purpose'' survey. Regarding sample size, we use as
reference, the best spectroscopic dataset of galaxy spectra at
z$\simlt$0.2, namely the ``classic'' Sloan Digital Sky Survey (SDSS), comprising
approximately 1 million optical spectra of galaxies brighter than
$r$$\sim$18\,AB \citep[e.g.,][]{SDSS}. The problems facing galaxy formation studies are not as
``clean'' as, for instance, finding $w$ in a dark energy-dominated cosmology,
or water vapour in an 
exoplanet. Galaxy formation is a highly complex field, involving a
large set of physical mechanisms. Such complex questions need large
datasets to be able to probe in detail variations of the observables
with respect to properties such as the stellar mass, size or
morphology of the galaxy under consideration, the mass of its host
halo, the potential nuclear activity (ongoing or recent), the presence
of infall/outflows, or tidal interactions from nearby
interlopers. Therefore, {\sl it is necessary to probe this
multi-parameter space in sufficient depth to understand in detail
the role of the mechanisms driving galaxy formation}. This is where a
large multiplex mission such as {\tt Chronos} exceeds the specifications
of future large facilities such as {\sl JWST} or {\sl ELT}. Although
JWST/NIRSpec will obviously have the capability of observing deep NIR
spectra of distant galaxies, its small field of view, lower multiplex
and oversubscription -- across a wide range of disciplines -- will
allow such a powerful telescope to gather, at most, $\sim$1,000 galaxy
spectra at similar spectral resolution, within the remit of this science
case \citep{Rieke:19}. Doubtlessly, it will
help tackle the science drivers listed above. However, such a small
sample will always lead to the question of whether the observed
sources are representative. Furthermore, if one wants to explore the
effect of one of the parameters/observables listed above, it will be
necessary to divide the sample accordingly. As an example, studies of
environment-related processes done at lower redshift with SDSS or
GAMA, work with samples between 10 and 100 times larger than the
potential output of JWST. {\sl{\tt Chronos} should be considered a
  successor to JWST and ELT-class facilities in galaxy formation
  studies.}

\subsection{Why the proposed spectral mode?}
\label{SS:WhySpec}

The table in Fig.~\ref{fig:DMD} (right) shows the overal properties of
the proposed survey. Choosing a wider wavelength coverage would enable
us to target additional spectral features. In principle, it would be
possible to extend the coverage to $K$ band. Note that the targeted
spectral coverage is suitable for the analysis of the feature-rich
region around the 4000\AA\ break in the ``cosmic-noon'' sample. Those
features will be good enough to determine accurate kinematics, stellar
population properties (age and chemical composition) and gas
parameters. Extending the data, e.g. to 3\,$\mu$m would add H$\alpha$
at the highest redshifts of the ``cosmic noon'' sample (z$\sim$4),
which will obviously increase the science return, as the H$\alpha$
region will allow us to improve on estimates of star formation rates,
or characterize in more detail the ionization state of the gas. In the
``cosmic dawn'' sample, the key region is the Ly$\alpha$ interval,
fully covered at the required redshifts by the proposed wavelength
range, so a limited extended spectral coverage in this sample is not
so beneficial.

However, we emphasize that a significant increase in wavelength
coverage at fixed spectral resolution could make the spectrograph
prohibitively expensive in a high multiplex system such as {\tt
  Chronos}.  The main science drivers cannot be fulfilled at lower
spectral resolution, and a lower multiplex will make the proposed
survey size unfeasible within a 5 year mission concept, so a wider
spectral coverage should not be the major direction to improve on this
concept. Note also that {\tt Chronos} will operate with ultra-faint
sources. At longer wavelengths, the thermal background of the
telescope will impose additional costly solutions to keep the
background at acceptable levels.

\subsection{Comparison with current and future projects}
\label{SS:Sci_Comp}

{\tt Chronos} will play a unique role in the landscape of future
near-infrared spectroscopic surveys. The MOONS multi-fiber
spectrograph at the ESO VLT will have a multiplex of about 1000 fibers
over a field of view (FoV) of 500\,arcmin$^2$, and will cover
0.6--1.8\,$\mu$m at R$\sim$5000 \citep{MOONS}. The Subaru Prime Focus
Spectrograph (PFS) will use up to 2400 fibers over 1.3\,deg$^2$, and
will cover 0.38-1.26\,$\mu$m at R$\sim$4300 in the near-IR ($\lambda
>$0.94\,$\mu$m) \citep{PFS}. The Maunakea Spectroscopic Explorer (MSE)
is planned as a dedicated 10m-class telescope with a high multiplex
(2,000--3,000) spectrograph that will work at low and high spectral
resolution \citep{MSE}.
However, the data taken with these
promising facilities will be inevitably affected by the atmosphere
(opaque spectral windows, telluric absorption lines, OH emission line
forest, high sky background) which will severely limit the
sensitivity, the quality of the spectra and the capability to observe
the continuum of faint objects. Having both continuum and absorption
information in galaxy spectra over a wide spectral window allows us
to break the degeneracies that entangle the properties of the stellar
populations. In the case of space-based facilities,
{\sl JWST} will have a very small survey efficiency due to its small
FoV and is therefore expected to play a complementary role in the
detailed study of small samples of objects. {\sl Euclid} and {\sl
  WFIRST} will survey very wide sky areas (15,000 and 5000\,deg$^2$,
respectively) in the near-infrared ($\sim$1--2\,$\mu$m), but the
spectroscopy will be slitless and with low resolution (R$<$500). This
makes {\sl Euclid} and {\sl WFIRST} powerful missions for redshift
surveys based on fairly bright emission lines, but less suitable for
detailed spectroscopic studies. For these reasons, {\tt Chronos} will
play a unique and unprecedented role thanks to its uninterrupted and
wide near-IR spectral range (rest-frame optical for z$>$1 objects),
extremely high sensitivity due to the low background, capability to
detect the continuum down to H=24-26AB, high S/N ratio suitable to
perform astrophysical and evolutionary studies, very high
multiplexing, wide sky coverage and large (SDSS-like) samples of
objects.

\subsection{Star formation history of galaxies}
\label{SS:Sci_SFH}

The stellar component in a galaxy is made up of a complex mixture of
ages and chemical composition, reflecting its past formation
history. For instance, galaxies that underwent recent episodes of star
formation will include a young stellar component, characterised
by strong Balmer absorption \citep[e.g.][]{Wild:09}; an efficient process
of gas and metal outflows will be reflected in the chemical
composition, targeted through metallicity-sensitive spectral indices
such as Mgb and $\langle$Fe$\rangle$ \citep[e.g.][]{SCT:00}; abundance variations between different
chemical elements, such as [Mg/Fe], map the efficiency of star
formation \citep[e.g.][]{Thomas:05};
variations in the stellar initial mass function (a fundamental component
of any galaxy formation model) can be constrained through the
analysis of gravity-sensitive indices \citep[e.g.][]{Ferreras:13,AH:18}.

\begingroup
\centering
\includegraphics[width=85mm]{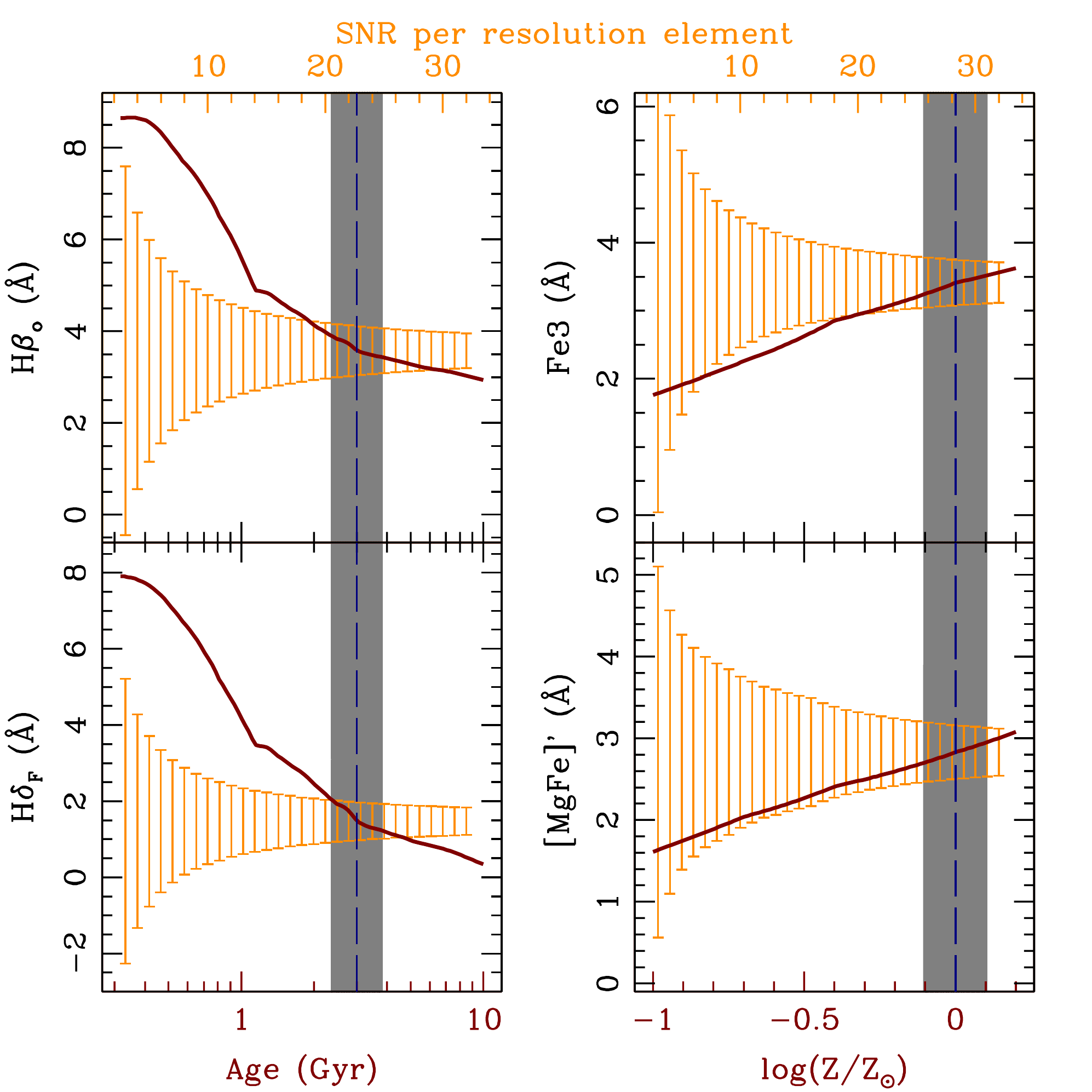}
\captionof{figure}{The red lines are model predictions from
  \citep{BC:03} for two age-sensitive ({\sl left}) and two metallicity
  sensitive ({\sl right}) line strengths for a galaxy with velocity
  dispersion $\sigma=$200\,km\,s$^{-1}$, as a function of age and
  metallicity, respectively (the bottom axes show the age and
  metallicity ranges). The orange lines are the estimated
  measurements, along with a 1\,$\sigma$ error bar, given as a
  function of S/N (shown in the top axes).  The simulated data
  correspond to a population at solar metallicity and age 3\,Gyr,
  marked with vertical dashed blue lines, along with a $\pm$0.1\,dex
  interval in grey.}
\label{fig:SNR}
\vskip+0.2truecm
\endgroup

The stellar component of a galaxy encodes a
fossil record of its evolution.  In contrast, the gaseous component
gives a snapshot of the ``ongoing'' processes. The analysis of the
unresolved stellar populations in distant galaxies is tackled
through targeted line strengths and spectral fitting, by comparing
high-quality spectroscopic data with the latest stellar population
synthesis models \citep[e.g.][]{Vazdekis:12,Vazdekis:15}.  Such
methods have been very successful at understanding the formation
history of low redshift galaxies by use of spectra from the Sloan
Digital Sky Survey \citep[e.g.][]{Gallazzi:05}. Similar type of
studies at high redshift are fraught with the difficulties of dealing
with very faint sources, in an observer frame (NIR) where the complex
and highly variable airglow and telluric absorption makes ground-based
observations tremendously challenging. Figure~\ref{fig:SNR} shows a
test with synthetic spectra of the S/N level required to constrain
stellar population parameters from a set of line strengths. For a
0.1\,dex (statistical) accuracy in log(Age) or log(Z/Z$_\odot$),
typical values of S/N of $\sim$10--20 per resolution element are required
{\sl in the continuum}. This is a challenging target for galaxies at
z$\sim 2-3$, given the faint flux levels in the continuum shown in
Fig.~\ref{fig:Hlimit}.

\subsection{The role of AGN}

Studies of the past star formation histories of galaxies
(\S\S\ref{SS:Sci_SFH}) need to be compared with diagnostics of AGN
activity, to understand the connection between galaxy growth and that
of the central SMBH. Such studies are based on emission line diagrams
\citep[e.g.][]{BPT} that trace the ionisation state of the
interstellar medium. The requirements with regards to the S/N and
spectral resolution are similar to the limits imposed by the analysis
of stellar populations, although we note that emission line
constraints will be less stringent, in general, to those in the
continuum. At high enough S/N, it may be possible to separate the
central component (dominated by the AGN) from the bulk of the
galaxy. As reference, a 0.1\,arcsec resolution element maps into a
projected physical distance of 0.8--1\,kpc at z$\sim$1--3.

\subsection{Environment and Merger history of galaxies}
\label{SS:Sci_Env}
Large spectroscopic redshift surveys are needed to characterize the
environment of galaxies in detail
\citep[e.g.][]{Yang:07,Robotham:11}. A mass-limited complete survey
will allow us to probe the merging history of galaxies, either from
the study of dynamically close pairs \citep[e.g.][]{CLS:12,SH4} or
through morphological studies \citep[e.g.][]{Lofthouse:16}.
 
Although deep NIR imaging surveys will be available at the time of a
potential L4 mission, there will not be a comprehensive counterpart of
spectroscopic observations, except for reduced sets of galaxies
($\simlt$1,000) observed by {\sl JWST}, or {\sl E-ELT}-like telescopes
from the ground. In order to beat cosmic variance it is necessary to
obtain spectroscopic redshifts covering large enough volumes. As a
rough estimate, we use the state-of-the-art Sloan Digital Sky Survey
as reference. The original low-redshift dataset, limited to
r$<17.7${\sc AB}, can be considered ``complete'' out to redshift
z$\simlt$0.2, covering a comoving volume of $5.5\times
10^{-5}$\,Gpc$^3$ per square degree. At the peak of galaxy formation
activity, z$\sim 1-3$, the equivalent volume is $0.02$\,Gpc$^3$ per
square degree. Since the SDSS footprint extends over $\sim
10^4$\,deg$^2$ on the sky, a similar comoving volume will be probed by
{\tt Chronos} if covering 30\,deg$^2$. Although a detailed analysis is
beyond the scope of this proposal, it may be advisable to opt for a
tiered survey, from shallower samples (H$<$24{\sc AB}) over
100\,deg$^2$ to deeper regions, covering $\sim 5-10$\,deg$^2$ at
H$<$26{\sc AB}.

\subsection{Gas and stellar kinematics and chemistry}

The emission and absorption line positions and shapes are a valuable
tool to study the kinematics and chemical composition of the stellar
and gaseous components.  Through high volumes of high S/N data with
high enough spectral resolution, it will be possible to trace stellar
kinematics and the mechanisms of gas outflows and stellar
feedback. Moreover, information such as the velocity dispersion
or the spin parameter can be used to constrain the
properties of the dark matter halos hosting galaxies at z$\sim$1-3
\citep[e.g.][]{Burkert:16,Wuyts:16}. More detailed analyses can be
gathered by integral field units, where the spectra of different
regions of the galaxy are extracted separately. Such instruments have
facilitated detailed analyses of the stellar and gaseous components in
nearby \citep[e.g. ATLAS$^{\rm 3D}$][]{Cap:11} and distant
\citep[e.g. KMOS$^{\rm 3D}$][]{Wisnioski:15} galaxies.  Due to the
faintness of the sources and the need for a high multiplex system
covering a wide field of view, we would, in principle, decide against
an IFU-based instrument, although this issue would be an important one
to tackle during the definition phase (see \S\ref{sec:Measure}).  Also
note that at the redshifts probed, the (spatial) resolving power is
rather limited, expecting a resolution -- measured as a physical
projected distance -- around 1\,kpc at z$\sim$1--3. The high S/N
requirements of the previous cases align with this one, but on the
issue of kinematics, a slightly higher spectral resolution may be
desired. Note also that at high resolution, high S/N spectra may be
used to disentangle different components \citep[such as bulge and
  disk,][]{Ocvirk:06}.

\subsection{Observations of galaxies in the high redshift Universe}

Deep NIR spectroscopy from space is the only way to confirm the
continuum break at 1216\AA\ in the high redshift Universe
(z$\simgt$6). Ground-based instruments only detect these objects when
they have strong Ly-$\alpha$ lines in clean regions of the night sky
spectrum. This line can be scattered by neutral intergalactic gas, and
is likely to be weaker at such redshifts. The goal of achieving enough
S/N in the continuum is important to properly characterize the
properties of the underlying stellar populations, something that could
be done with individual galaxies in the deep survey, and with stacked
subsamples in the wide survey.

\end{multicols}

\section{Measurement Concept}
\label{sec:Measure}

\begin{multicols}{2} 

\subsection{Reconfigurable Focal Planes for Space Applications}
\label{SS:ReconfFP}

{\sl JWST} will be the first astronomy mission to have a true multi-object
spectroscopic capability via the micro-shutter arrays in the NIRSpec
instrument, which can observe up to ~100 sources simultaneously over a
field-of-view around $3^\prime\times 3^\prime$ \citep{Li:07}. Scaling
this technology to the field sizes and multiplex advantage required
for the next generation of space-based spectroscopic survey
instruments is not straightforward however, and will likely require a
new approach. There are currently three technologies which show
promise in this area.

\begin{figure*}
\centering
\raisebox{-0.55\height}{\includegraphics[width=70mm]{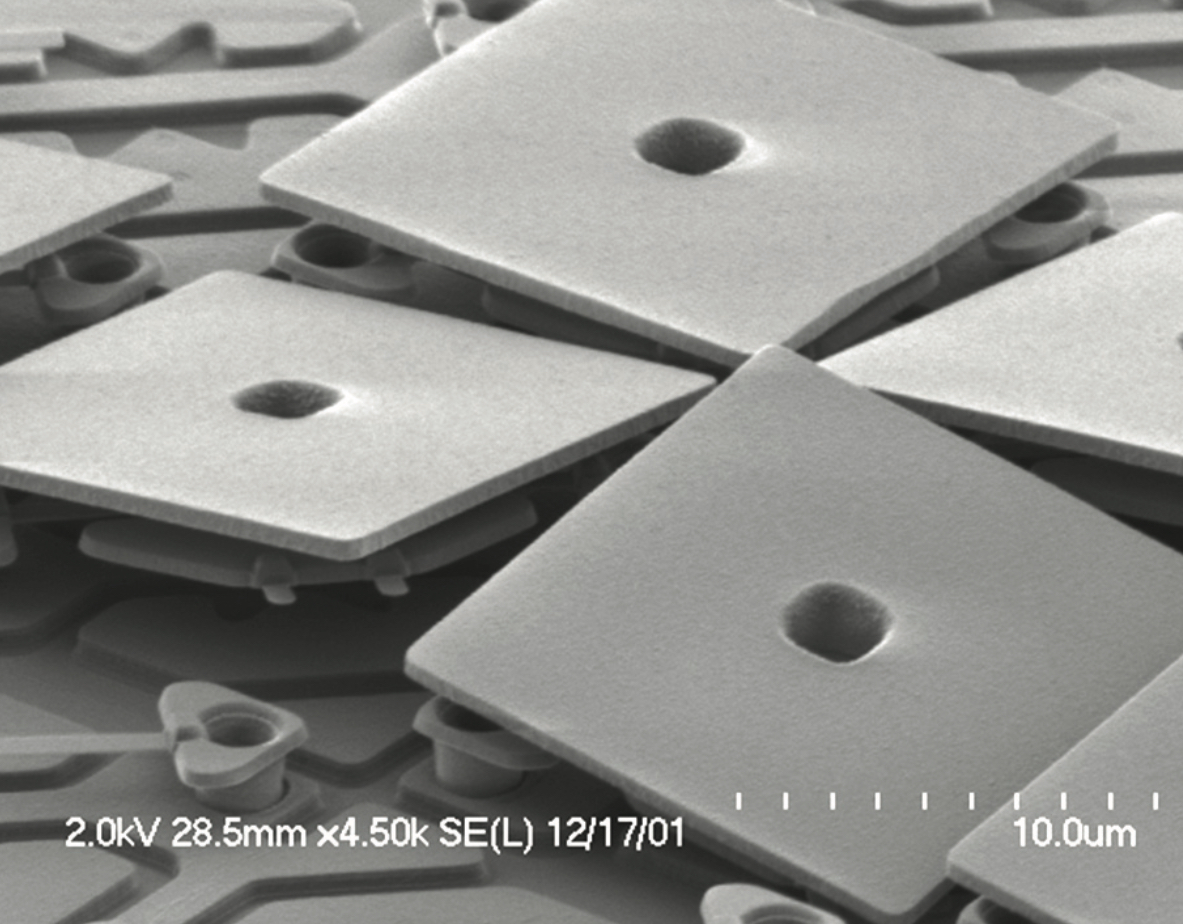}}
\hspace{0.5truecm}
\centering 
{\parbox{105mm}{\begin{tcolorbox}[tab1,tabularx={X|C{21mm}|C{21mm}},title=Summary of {\tt Chronos} survey specifications,boxrule=0.5pt]
Spectral range  & \multicolumn{2}{c}{0.8--2$\mu$m}\\ \hline
Spectral resolution & \multicolumn{2}{c}{1500-3000}\\ \hline
Target Multiplex & \multicolumn{2}{c}{$\simgt 5000$}\\ \hline
Field of View (deg$^2$) & \multicolumn{2}{c}{$\simgt$0.2}\\ \hline\hline
SURVEY & Wide & Deep\\ \hline
Sensitivity (@S/N=20) & H=24{\sc AB} & H=26{\sc AB}\\ \hline
Line Sensitivity (@5$\sigma$) & $5\times 10^{-19}$cgs  & $8\times 10^{-20}$cgs\\ \hline
Galaxy density (z=1--3, deg$^{-2}$) & $4.8\times 10^4$ & $1.2\times 10^5$\\ \hline
Coverage (deg$^2$) & 100 & 10\\ \hline
\end{tcolorbox}
}}
\captionof{figure}{{\sl Left:} Photomicrograph of tilted DMD
  micromirrors. The neighbouring mirrors have been removed to reveal
  the substructure (courtesy ASME/Texas Instruments). {\sl Right:}
  General specifications of the proposed survey.}
\label{fig:DMD}
\end{figure*}

\begin{enumerate}
\item\underline{Digital Micromirror Arrays (DMDs):} Digital micromirror
technology was developed in the 1990s by Texas Instruments for use in
light projection systems (see Fig.~\ref{fig:DMD}). The current
state-of-the-art is 2kx1k devices with 13\,$\mu$m pixels but larger
format devices (up to 16 million pixels) are under development
(c.f. 62,000 micro-shutters in {\sl JWST} NIRSpec). These devices are
also well matched in pixel size to the focal planes of small to medium
size telescopes.  DMDs were first proposed for the ESA M-class SPACE
mission concept \citep{CI:09} which later evolved into the {\sl
  Euclid} mission. The primary technological challenges in exploiting
DMDs in space are: (i) developing radiation-hard electronics to drive
the DMDs (the MEMS technology used in the mirrors themselves are not
susceptible to damage except by extreme micrometeorite events), (ii)
demonstrating reliable operation at cryogenic temperatures as required
for observations at near-infrared wavelengths, (iii) modifying the
visible-light windows on commercial devices to allow extended
operation into the near-infrared, (iv) improving the
contrast/scattered light for bright objects. Preliminary work has been
undertaken during studies for SPACE/{\sl Euclid} and elsewhere
\citep{ZA:11,ZA:17} but further work is required to raise the TRL.

\item\underline{Reconfigurable Slits:} A near-infrared multi-object
spectroscopy target selection system which has been successfully
deployed on the ground-based MOSFIRE instrument at Keck is the
Configurable Slit Unit (CSU) \citep{Span07}. This is a form of
micro-mechanical system which employs voice-coil actuated
``inch-worm'' motors to position up to 46 slitlets (each 5\,mm long)
in a $4^\prime\times 4^\prime$ field.  This technology has been proven
to be reliable in cryogenic operation (at 120K) over several
years. Whilst the 1-D motion of the slits in the focal plane reduces
target acquisition efficiency somewhat, the contiguous slits allow for
improved sky-subtraction compared to devices (like the DMD) where the
local sky is obtained via separate apertures. The primary technology
challenges of adopting this technology for space applications would be
the substantial miniaturization required and increasing the multiplex
gain by 1 or 2 orders of magnitude, possibly using a piezo-electric
drive system.

\item\underline{Liquid Crystal Masks:} Liquid crystal (LC) masks are
widely used as spatial light modulators in a number of laboratory
applications. Whilst fundamentally relying on the ability of
polarizing crystals to transmit or block linearly polarized light,
they can be made to work more efficiently on unpolarized light using
polymer dispersed liquid crystals (PDLCs). By combining an optically
active material with an appropriate electrode structure,
reconfigurable masks can be obtained which only transmit light in a
specific spatial pattern \citep{WS92}. Devices in formats up to
1024x768 are commercially available with up to 36\,$\mu$m pixels. The
primary technological challenges would be: (i) operation efficiency
(switching times) at low temperatures due to the properties of the LC
medium, (ii) contrast limits (many commercial devices have contrast
ratios <100:1), (iii) limitations on bandwidth due to the chromatic
properties of LCs, (iv) limitations due to non-orthogonal
illumination.
\end{enumerate}

In addition to the above reconfigurable focal plane solutions, which
segment the focal plane spatially according to preselected target
positions, an alternative approach is to select targets from a fixed
grid of sub-areas across the focal plane (one target per sub-area)
using a ``beam-steering'' approach. Many of these rely on similar
underlying technologies to those discussed above (i.e. MEMS and/or
variable prisms) and should be explored in the context of specific
mission requirements. They are particularly suitable to selecting
targets for spatially-resolved (``integral field'') spectroscopic
studies.

\subsection{Large format Integral Field Units}

A complementary approach to massively multiplexed spectroscopy when
the target densities are high enough, is to use some form of integral
field spectroscopy (IFS) which delivers a full spectral datacube for a
contiguous region of sky. The IFS approach also opens up a large
serendipity space since no imaging surveys are required to pre-select
targets.

Integral field units have been widely used on ground-based telescopes
\citep[e.g.][]{AllS:06} and a small-format device ($30\times 30$
spatial pixels) will be launched on {\sl JWST} as one of the observing modes
for the NIRSpec spectrograph \citep{Birkman:14}. Integral field
systems can be realized using a number of techniques but the favoured
approach for space infrared systems is the diamond-machined image
slicer \citep{Lobb:08} which can take advantage of monolithic
manufacturing methods and a robust thermal design approach. Technology
developments would be required to develop wide-field integral field
systems for space applications, but the generic approach using a
``field-splitter'' front-end optic to feed multiple sub-systems is
well-understood from ground-based instruments
\citep[e.g.][]{Pares:12}. Mass, power and data rate budgets remain to
be explored but will be common to all wide-field spectroscopic
facilities.

\subsection{NIR detector technology}

A successful outcome of a survey such as {\tt Chronos} also rests on highly
efficient NIR detector technology, with minimal noise and
well-understood systematics. The survey operates at a very low-photon
regime, where it is essential to control the noise sources, and to
understand in detail the response of the detector. For instance,
cross-talk and persistence are substantial problems that can hinder
the observations, and need to be characterised in exquisite detail.

The best available technology for this science case involves
HgCdTe-based detectors, where the spectral range can be optimised by
the choice of the ratio of Hg to Cd, that modifies the band gap
between 0.1 and 1.5\,eV. As of today, US-based companies can provide
4k$\times$4k HgCdTe arrays with high enough TRL for a space mission
\citep[e.g. Teledyne,][]{Teledyne}.  However, given the long
timescales expected for a potential mission, and aligned with ESA's
investment in NIR detector technology development \citep{Nelms:16}, it
would be desirable to involve European groups (such as CEA-LETI, Selex
ES or Caeleste) in the development of ultra-sensitive NIR detectors
and the associated electronics. New HgCdTe-based technology with
avalanche photodiodes, developed by Selex appears quite promising for
astrophysics applications \citep{Saphira}.

\subsection{Photonics-based approach}

An alternative approach to the traditional spectrograph design is to
adopt a photonics-based instrument, creating the equivalent version of
an integrated circuit in electronics.  Astrophotonics has produced
several revolutionary technologies that are changing the way we think
of conventional astronomical instrumentation. In particular, the
invention of the photonic lantern \citep{LBB:05} allows us to reformat
the input to any instrument into a diffraction-limited output. As was
first described in \citet{JBH:06} and \citet{JBH:10}, this means that,
in principle, {\sl any spectrograph operating at any resolving power
  can be designed to fit within a shoebox}. These authors refer to
this as the photonic integrated multimode microspectrograph (PIMMS)
concept and it has been demonstrated at the telescope and in space
(see Fig.~\ref{fig:pimms}). Suitable optical designs are presented in
\citet{RB:12}.  Presently, the main limitation is that the ideal
detector has yet to be realized, although discussions are ongoing with
detector companies. This technology is ideally suited to optical and
infrared spectroscopy, and may overcome the technological challenges faced
by conventional spectrograph designs within ESAs Voyage 2035-2050
long-term plan.

\begingroup
\vskip+0.2truecm
\centering
\includegraphics[width=80mm]{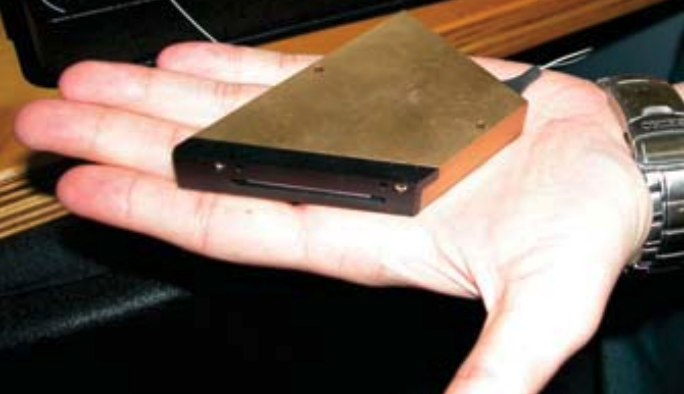}
\captionof{figure}{As seen above, the PIMMS concept has been
  demonstrated at the telescope \citep{Cvet:12}, on a balloon
  (2012) and onboard the Inspire cubesat (2017-8) flown by the
  University of Sydney \citep{Cairns:19}.}
\label{fig:pimms}
\vskip+0.2truecm
\endgroup

\end{multicols}

\clearpage

\begin{multicols}{2} 


\setlength{\bibsep}{2.0pt}



\end{multicols}

\end{document}